\documentclass[%
 reprint,
nofootinbib,
 amsmath,amssymb,
 aps,
 prd,
]{revtex4-2}

\usepackage[normalem]{ulem}
\usepackage{multirow}
\usepackage{graphicx}% Include figure files
\usepackage{dcolumn}% Align table columns on decimal point
\usepackage{bm}% bold math
\usepackage[dvipsnames]{xcolor}
\usepackage{hyperref}% add hypertext capabilities
\usepackage{bookmark}
\usepackage{enumerate}
\usepackage{CJKutf8} 
\newcommand{\UM}{Department of Physics, University of Michigan, 450 Church St, Ann Arbor, MI 48109, USA}
\newcommand{\LCTP}{Leinweber Center for Theoretical Physics, 450 Church St, Ann Arbor, MI 48109, USA}

\begin{document}

\preprint{APS/123-QED}

\title{Design and optimization of neural networks for multifidelity cosmological emulation}% Force line breaks with \\

% \author{\begin{CJK}{UTF8}{gbsn}Yanhui Yang (\CJKfamily{gbsn}杨焱辉)$^1$\end{CJK}}
\author{Yanhui Yang (\begin{CJK}{UTF8}{gbsn}杨焱辉\end{CJK})$^1$}
 \email{yyang440@ucr.edu} 
\author{Simeon Bird$^1$}%
 \email{sbird@ucr.edu}
% \author{\begin{CJK}{UTF8}{bsmi}Ming-Feng Ho (何銘峰)$^{1,2,3}$\end{CJK}}
% \author{Ming-Feng Ho (\begin{CJK*}{UTF8}{bsmi}何銘峰\end{CJK*})}
% test without CJK
\author{Ming-Feng Ho (\begin{CJK*}{UTF8}{bsmi}何銘峰\end{CJK*})$^{1,2,3}$}
 \
\author{Mahdi Qezlou$^4$}

\affiliation{%
$^1$Department of Physics \& Astronomy, University of California, Riverside, 900 University Ave., Riverside, CA 92521, USA
}%
\affiliation{$^2$\UM}
\affiliation{$^3$\LCTP}
\affiliation{$^4$The University of Texas at Austin, 2515 Speedway Boulevard, Stop C1400, Austin, TX 78712, USA
}

% \collaboration{MUSO Collaboration}%\noaffiliation

%  \homepage{http://www.Second.institution.edu/~Charlie.Author}
% \affiliation{
%  Second institution and/or address\\
%  This line break forced% with \\
% }%
% \affiliation{
%  Third institution, the second for Charlie Author
% \affiliation{%
%  Authors' institution and/or address\\
%  This line break forced with \textbackslash\textbackslash
% }%

% \collaboration{CLEO Collaboration}%\noaffiliation

\date{\today; published in Phys. Rev. D on February 9, 2026}% It is always \today, today,

\begin{abstract}

Accurate and efficient simulation-based emulators are essential for interpreting cosmological survey data down to nonlinear scales. Multifidelity emulation techniques reduce simulation costs by combining high- and low-fidelity data, but traditional regression methods such as Gaussian processes struggle with scalability in sample size and dimensionality. In this work, we present {\scriptsize T2N-MusE}, a neural network framework characterized by (i) a novel 2-step multifidelity architecture, (ii) a 2-stage Bayesian hyperparameter optimization, (iii) a 2-phase $k$-fold training strategy, and (iv) a per-$z$ principal component analysis strategy. We apply {\scriptsize T2N-MusE} to selected data from the {\scriptsize Goku} simulation suite, covering a 10-dimensional cosmological parameter space, and build emulators for the matter power spectrum over a range of redshifts with different configurations. We find the emulators outperform our earlier Gaussian process models significantly and demonstrate that each of these techniques is efficient in training neural networks or/and effective in improving generalization accuracy. We observe a reduction in the mean error by more than a factor of five and in the worst-case error by approximately a factor of eight in leave-one-out cross-validation, relative to previous work. This framework has been used to build the most powerful emulator for the matter power spectrum, {\scriptsize GokuNEmu}, and will also be used to construct emulators for other statistics in future.
%, e.g., the Lyman-$\alpha$ forest flux power spectrum.

% \begin{description}
% Secondary publications and information retrieval purposes.
% You may use the \texttt{description} environment to structure your abstract;
% use the optional argument of the \verb+\item+ command to give the category of each item. 
% \end{description}
\end{abstract}
    
\maketitle

%\tableofcontents

\section{\label{sec:intro}Introduction}

Cosmological surveys such as the Dark Energy Spectroscopic Instrument (DESI)~\cite{DESICollaboration2016}, Legacy Survey of Space and Time (LSST)~\cite{LSC2009}, \textit{Euclid}~\cite{Laureijs2011}, the Nancy Grace Roman Space Telescope~\cite{Akeson2019}, the China Space Station Telescope (CSST)~\cite{Gong2019}, and the Prime Focus Spectrograph (PFS) on the Subaru Telescope~\cite{Takada2014} will enable precise measurements of the galaxy power spectrum, as well as the weak lensing shear field.
These measurements will be used to constrain cosmological models motivated by unresolved fundamental physics questions.

Interpreting the data and inferring cosmological parameters requires making predictions for the matter field or a summary statistic, such as the matter power spectrum, and using Bayesian methods. A naive inference run may require $10^6$--$10^7$ matter power spectrum evaluations at different cosmological parameters, which would be computationally expensive.

Emulation replaces intensive numerical computation for every likelihood evaluation by the evaluation of a cheap pre-trained surrogate model. For instance, emulators have been widely used to replace the Boltzmann codes in cosmological inference~\cite{Auld2007,Auld2008,Arico2021a,Spurio2022,Nygaard2023,Gunther2022,Bonici2024,Bonici2025}. Emulators based on $N$-body simulations are needed to interpret observations on nonlinear scales, $k \gtrsim 0.1h$/Mpc. There have been several such cosmological emulators, e.g., {\scriptsize  FrankenEmu}~\cite{Heitmann2009,Heitmann2010,Heitmann2013}, the emulators of the Aemulus project~\cite{DeRose2019,McClintock2019,Zhai2019}, {\scriptsize  NGenHalofit}~\cite{Smith2019}, the {\scriptsize  dark quest} emulator~\cite{Nishimichi2019}, {\scriptsize  BE-HaPPY}~\cite{Valcin2019}, the baryonification emulator of the BACCO project~\cite{Arico2021}, the emulators built on the \textsc{Quijote} simulations~\cite{Villaescusa2020}, the emulators based on the \textit{Mira-Titan Universe} suite~\cite{Heitmann2016,Lawrence2017,Bocquet2020,Moran2023,Kwan2023}, the {\scriptsize E-MANTIS} emulator~\cite{Casares2024}, {\scriptsize  EuclidEmulator}~\cite{Knabenhans2019,Knabenhans2021}, {\scriptsize CSST Emulator}~\cite{Chen2025} and {\scriptsize GokuEmu}~\cite{Yang2025}.
These emulators are able to predict summary statistics within their parameter space with orders of magnitude lower computational costs than full simulations.

% parameter coverage
There are several well-motivated extensions of the standard cosmological model which are constrained by current and future surveys. However, including these extensions in emulators is challenging due to the high dimensionality of the parameter space that necessitates a large number of computationally expensive samples. Multifidelity (MF) techniques have been developed to reduce the computational cost of building emulators, e.g., {\scriptsize MFEmulator}~\cite{Ho2022} and {\scriptsize MF-Box}~\cite{Ho2023}. Ref.~\cite{Yang2025} built {\scriptsize  GokuEmu}~\cite{Yang2025}, an emulator for the matter power spectrum, which expanded the parameter space to 10 dimensions for the first time, taking into account dynamical dark energy, massive neutrinos, the effective number of ultrarelativistic neutrinos and the running of the primordial spectral index. This was achieved by using {\scriptsize MF-Box}, which combines simulations with different box sizes and particle loads, at a computational cost 94\% less than single-fidelity approaches.

Despite the success of {\scriptsize MF-Box} in reducing the computational cost of producing the training data (simulations), the regression technique used, Gaussian process (GP) regression, still suffers from the curse of dimensionality. The computational complexity of GP regression scales poorly (cubically) with sample size (see Chapter 8 of Ref.~\cite{2006gpml.book.....R} or Chapter 9 of Ref.~\cite{garnett_bayesoptbook_2023}). This in turn leads to lengthy prediction and training times, as well as increased memory usage. GP regression struggles to satisfy our need for next-generation cosmological emulators, which would ideally become yet more complex, including non-standard dark matter models or baryonic physics.

Neural networks (NNs) have been used in emulators. For example, Ref.~\cite{Cabayol-Garcia2023} built an NN emulator for the Lyman-$\alpha$ forest 1D flux power spectrum, Ref.~\cite{Diao2025} constructed an MF emulator for large-scale 21\,cm lightcone images using generative adversarial networks, and Ref.~\cite{Zhang2025} trained models for gravitational waves using NNs. NNs are suitable for larger data sets, given that they typically scale linearly or sublinearly with sample size (see, e.g., Ref.~\cite{Hestness:2017arXiv171200409H}). They are also more efficient in inference time and memory usage. In addition, Ref.~\cite{Guo2022} showed that NN MF regression can outperform GP regression in terms of accuracy in some cases and suggested that a high-dimensional parameter space would prevent GP regression from being effective.
 
In this work, we develop the ``Triple-2'' neural network framework for multifidelity cosmological emulation ({\scriptsize T2N-MusE}), characterized by a ``2-step'' MF architecture, a ``2-stage'' hyperparameter optimization process, and a ``2-phase'' $k$-fold training strategy. Compared to Ref.~\cite{Guo2022}, we have made several improvements. We introduce a modified ``2-step'' MF architecture, which turns out to be more suitable in the context of cosmological emulation than the original ``2-step'' architecture. The ``2-stage'' hyperparameter optimization process and ``2-phase'' training strategy further improve the emulation performance. In addition, we propose a per-redshift data compression strategy to further boost the emulator's accuracy. We test the performance of {\scriptsize T2N-MusE} on selected data from the {\scriptsize Goku} simulation suite~\cite{Yang2025}, demonstrating the efficacy of these training strategies.

% paper structure
We organize this paper as follows. Sec.~\ref{sec:methods} introduces the cosmological simulation data used in this study (Sec.~\ref{sec:sim_data}), the MF architectures of the neural networks (Sec.~\ref{sec:arch}), the workflow of training the neural networks (Sec.~\ref{sec:workflow}), the comparative study we design to evaluate the performance of different choices of architectures and strategies for data compression and optimization of NNs (Sec.~\ref{sec:compare}). In Sec.~\ref{sec:results}, we present the results of the comparative study, showing the effects of different approaches on the emulation performance. Finally, we conclude in Sec.~\ref{sec:concl}.

\section{\label{sec:methods}Methods}

\subsection{\label{sec:sim_data}Simulation Data}

We briefly recap the {\scriptsize Goku} simulation suite and the specific data we use in this work. This paper focuses on the machine learning techniques we have developed for building highly optimized emulation models. For more details on the simulation suite, please refer to Ref.~\cite{Yang2025}.

{\scriptsize Goku} is a suite of $N$-body simulations that covers 10 cosmological parameters, performed using the {\scriptsize MP-Gadget} code~\cite{Feng2018}. A relatively large number of low-fidelity (LF) simulations were sampled in the parameter space using a sliced Latin hypercube design~\cite{Qian2012}, and a small number of high-fidelity (HF) cosmologies were chosen from the LF cosmologies so as to optimize the available HF information. {\scriptsize Goku} includes two Latin hypercubes that cover different parameter boxes, {\scriptsize Goku-W} and {\scriptsize Goku-N}, with wide and narrow ranges of parameters, respectively. For convenience of testing the methods, only {\scriptsize Goku-W} will be used in this study. We note that the emulator trained on {\scriptsize Goku-W} exhibits larger generalization errors than that trained on {\scriptsize Goku-N}~\cite{Yang2025}, underscoring the need for improved modeling over this broader parameter space. Although there are two LF nodes, L1 and L2, we only use L2 in this study (hereafter, we refer to L2 as LF).\footnote{L2 corresponds to the $k$ range where emulation error dominates total uncertainty, making it more suitable for evaluating improvement from the applied techniques.} We have nevertheless verified that the techniques developed in this work can also improve models trained on L1 simulations (see Appendix~\ref{app:L1_optimal} for an NN emulator trained on the HF and L1 nodes of {\scriptsize Goku-W}). We summarize the LF and HF simulations in Table~\ref{tab:sims}.
The redshifts considered are $z = 0, 0.2, 0.5, 1, 2$ and $3$. The matter power spectra measured from these simulations, along with their cosmologies, are the data we use to train the neural networks.

Specifically, the input of the target model are the 10 cosmological parameters\footnote{In practice, they are normalized to $[-0.5, 0.5]$.}, i.e., the input vector $\mathbf{x} \in \mathbb{R}^{d_\mathrm{in}}$, where $d_\mathrm{in} = 10$. The output is the matter power spectrum at a series of $k$ modes and redshifts, i.e., the output vector
\begin{equation}
    \begin{split}
        \mathbf{y} =& \,[y(z_1, k_1), y(z_1, k_2), \ldots, y(z_1, k_{n_k}), \\
        & \,y(z_2, k_1), \ldots, y(z_2, k_{n_k}), \\
        & \,y(z_{n_z}, k_1), \ldots, y(z_{n_z}, k_{n_k})],
    \end{split}
    \label{eq:output}
\end{equation}
where $y(z_i, k_j) = \lg P(z_i, k_j)$ is the matter power spectrum in log space at redshift $z_i$ and wavenumber $k_j$, $n_z$ is the number of $z$ bins, and $n_k$ is the number of $k$ modes. The output vector $\mathbf{y} \in \mathbb{R}^{d_\mathrm{out}}$, where $d_\mathrm{out} = n_z \times n_k$. In our case with {\scriptsize Goku-W}, we have $n_z = 6$ and $n_k = 64$, hence $d_\mathrm{out} = 384$.

\begin{table}%[b]%The best place to locate the table environment is directly after its first reference in text
    \caption{\label{tab:sims}%
    Specifications and numbers of simulations in the {\scriptsize Goku-W} suite.
    }
    \begin{ruledtabular}
    \begin{tabular}{lccc}
    {Simulation}&
    {Box size}&
    {Particle}&
    {Number of}\\
    fidelity & ($\text{Mpc}/h$) & {\textrm{load}} & {simulations}\\
    \colrule
    HF & 1000 & $3000^3$& $n_\mathrm{H}= 21$\\  %averaged over 21/15 HF sims
    LF & 250 & $750^3$ & $n_\mathrm{L} = 564$\\
    \end{tabular}
    \end{ruledtabular}
\end{table}

\subsection{\label{sec:arch}Multifidelity Architectures}

Ref.~\cite{Guo2022} proposed a ``2-step'' architecture for NN MF regression, which consists of two NNs. A first NN is trained on the LF data, $\mathcal{T}^\mathrm{L}=\{(\mathbf{x}^{\mathrm{L},i}, \mathbf{y}^{\mathrm{L},i}): i=1,2,\ldots,n_\mathrm{L}\}$, to learn the LF function $f^L$, such that $\mathbf{y}^{\mathrm{L}} = f^\mathrm{L}(\mathbf{x})$. Then a second NN approximates the correlation between the LF and HF functions, $\mathcal{F}$, such that $\mathbf{y}^{\mathrm{H}} = \mathcal{F} (\mathbf{x}, \mathbf{y}^\mathrm{L})$, based on the input data $(\mathcal{X}^\mathrm{H}, f_\mathrm{NN}^\mathrm{L} (\mathcal{X}^\mathrm{H})) = \{(\mathbf{x}^{\mathrm{H},i}, f_\mathrm{NN}^\mathrm{L} (\mathbf{x}^{\mathrm{H},i})):i=1,2,\ldots,n_\mathrm{H}\}$ and the available HF output data $\mathcal{Y}^\mathrm{H} = \mathbf{y}^\mathrm{H}(\mathcal{X}^\mathrm{H})= \{\mathbf{y}^{\mathrm{H},i}: i=1,2,\ldots,n_\mathrm{H}\}$. Note that in our case, the HF cosmologies are a subset of the LF cosmologies, so we can replace $f_\mathrm{NN}^\mathrm{L} (\mathcal{X}^\mathrm{H})$ with $\mathbf{y}^\mathrm{L}(\mathcal{X}^\mathrm{H})$, such that the two NNs can be trained independently and simultaneously. While Ref.~\cite{Guo2022} restricts the second NN to be a shallow NN with only one hidden layer, we allow multiple hidden layers in the second NN to increase the flexibility of the model.

\begin{figure*}
    % \centering
    \includegraphics[width=0.84\linewidth,trim=360 710 680 90,clip]{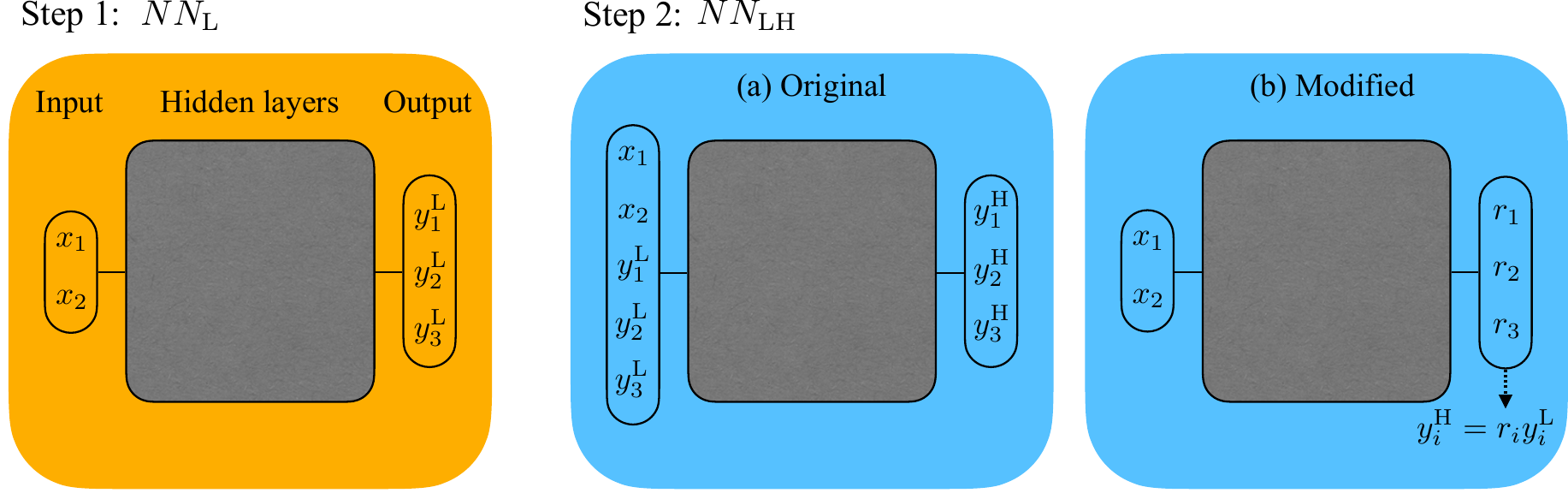}
    \caption{\label{fig:arch}Examples of the original and modified 2-step MF NN architectures. Both architectures have the same $NN_\mathrm{L}$ (step 1: the LF NN) but different $NN_\mathrm{LH}$ (step 2: the NN used to correct the LF output). The original $NN_\mathrm{LH}$ (a) approximates the correlation between the LF and HF functions, with $(\mathbf{x}, \mathbf{y}^\mathrm{L})$ as input and $\mathbf{y}^\mathrm{H}$ as output. The modified $NN_\mathrm{LH}$ (b) learns the mapping from $\mathbf{x}$ to the ratio of $\mathbf{y}^\mathrm{H}$ to $\mathbf{y}^\mathrm{L}$, $\mathbf{r} = \mathcal{G} (\mathbf{x})$, and the final HF output is the element-wise product of the LF output with the correction ratio $\mathbf{r}$.}
\end{figure*}

Figure~\ref{fig:arch} illustrates the original 2-step architecture with a simple example of 2D input and 3D output. Note that the input of $NN_\mathrm{LH}$ (the NN modeling LF-HF correlation), is a 5D vector, which is a concatenation of the LF output and the initial input vector.

We propose a modified 2-step architecture with the same $NN_\mathrm{L}$ but a different $NN_\mathrm{LH}$, illustrated in Fig.~\ref{fig:arch}. Instead of approximating the correlation between the LF and HF functions, the new $NN_\mathrm{LH}$ learns the ratio of $\mathbf{y}^\mathrm{H}$ to $\mathbf{y}^\mathrm{L}$, $\mathbf{r}$ with the component $r_i = y^\mathrm{H}_i / y^\mathrm{L}_i$ for $i=1,2,\ldots,d_\mathrm{out}$, as a function of the input vector $\mathbf{x}$, i.e., $\mathbf{r} = \mathcal{G} (\mathbf{x})$. The training data for $NN_\mathrm{LH}$ is $\mathcal{T}^\mathrm{H} = \{(\mathbf{x}^{\mathrm{H},i}, \mathbf{r}^{\mathrm{H},i}): i=1,2,\ldots,n_\mathrm{H}\}$, where $\mathbf{r}^{\mathrm{H},i} = \mathbf{y}^{\mathrm{H},i} \oslash f_\mathrm{NN}^\mathrm{L} (\mathbf{x}^{\mathrm{H},i})$. As before, we replace $f_\mathrm{NN}^\mathrm{L} (\mathbf{x}^{\mathrm{H},i})$ with $\mathbf{y}^\mathrm{L}(\mathbf{x}^{\mathrm{H},i})$. With the trained $NN_\mathrm{L}$ and $NN_\mathrm{LH}$, we can predict the HF output as $\mathbf{y}_\mathrm{NN}^\mathrm{H} = \mathcal{G}_\mathrm{NN}(\mathbf{x}) \odot f_\mathrm{NN}^\mathrm{L} (\mathbf{x})$.\footnote{
    The symbol $\odot$ denotes the element-wise multiplication, and $\oslash$ denotes the element-wise division.} Note that, for the matter power spectrum, the ratio is calculated in original space rather than log space.

The modified 2-step model significantly reduces the dimensionality of the input of $NN_\mathrm{LH}$, which is $d_\mathrm{in} + d_\mathrm{out}$ in the original architecture and $d_\mathrm{in}$ in the modified architecture. This is particularly important for high-dimensional output data, such as the matter power spectrum, where $d_\mathrm{out} \gg d_\mathrm{in}$.

We will test the performance of both architectures in our comparative study (Sec.~\ref{sec:compare}) and show the results in Sec.~\ref{sec:arch_result}.

\subsection{\label{sec:workflow}Neural Network Workflow}

\begin{figure}
    % \centering
    \includegraphics[width=0.85\linewidth,trim=720 490 770 95,clip]{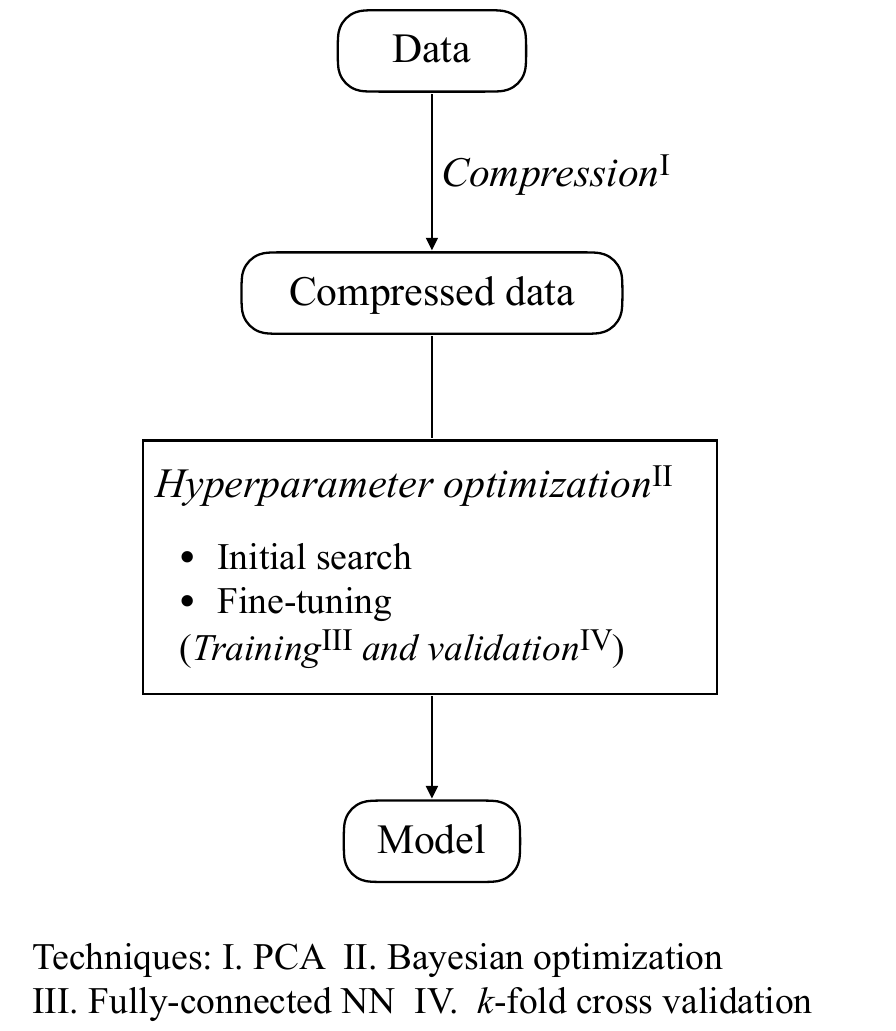}
    \caption{\label{fig:nn_workflow}Overview of the workflow of training a highly optimized NN. The workflow consists of three main steps: data compression, hyperparameter optimization, and training the final model. When optimizing the hyperparameters, a large space is explored in the initial search stage, and then a smaller space is searched in the second stage (fine-tuning). Each evaluation of the hyperparameters involves both training and validation of an NN.}
\end{figure}

We show a schematic of the training workflow for a highly optimized fully-connected neural network (FCNN) in Fig.~\ref{fig:nn_workflow}. First, we perform data compression to reduce the dimensionality of the output using principal component analysis (PCA). Then we explore the hyperparameter space of the neural networks through a two-stage Bayesian optimization process. In the first stage, we perform a coarse search over a large space of hyperparameters, and in the second stage, we perform a fine-tuning search over a narrower space. The bounds of the fine-tuning search are defined around the best-performing configurations found in the first stage. Each evaluation of the hyperparameters involves both training and validation of an NN.
Importantly, this pipeline is not limited to the multifidelity emulation context; it is broadly applicable to other tasks, including single-fidelity emulation and general regression problems involving high-dimensional outputs.

More details of each component of the workflow are given in their dedicated sections. See Sec.~\ref{sec:compress} for data compression, Sec.~\ref{sec:training} for neural network training, and Sec.~\ref{sec:hyper_opt} for hyperparameter optimization.

\subsubsection{\label{sec:compress}Data compression}

We use PCA to reduce the dimensionality of the output data. Two strategies are explored in this work: global PCA and per-redshift (hereafter, local) PCA. The former was adopted in some existing emulators, e.g., {\scriptsize EuclidEmulator2}~\cite{Knabenhans2021} and the {\scriptsize CSST Emulator}~\cite{Chen2025}. We propose the latter as a new approach to compress the output data, allowing a more flexible representation of the output data that may be better suited to the case where the redshift evolution of the output is nonlinear or complex.

In the global PCA approach, we perform PCA on all $k$ modes and redshifts together, and then each of the original output components can be expressed as a linear combination of the principal components (PCs), i.e.,
\begin{equation}
    y(z_i, k_j; \mathbf{x}) = \mu (z_i, k_j) + \sum_{l=1}^{n_\mathrm{PCA}} a_l(\mathbf{x}) {\phi}_l(z_i, k_j),
    \label{eq:pca_global}
\end{equation}
where $\mu (z_i, k_j)$ is the mean of the output data, $a_l(\mathbf{x})$ is the coefficient of the $l$th PC (i.e., eigenvector), $\phi_l(z_i, k_j)$ is the $l$th PC at redshift $z_i$ and wavenumber $k_j$, and $n_\mathrm{PCA}$ is the number of PCs. Then the compressed output can be expressed as
\begin{equation}
    \mathbf{y}_\mathrm{c}  = [a_1, a_2, \ldots, a_{n_\mathrm{PCA}}],
    \label{eq:pca_global_comp}
\end{equation}
reducing the dimensionality of the output from $n\times m$ [see Eq.~(\ref{eq:output})] to $d_\mathrm{out}^\mathrm{glob}=n_\mathrm{PCA}$.

In the local PCA, we perform PCA on each redshift separately. For redshift $z_i$, we have
\begin{equation}
    y(z_i, k_j; \mathbf{x}) = \mu^{i}(k_j) + \sum_{l=1}^{n_\mathrm{PCA}^{i}} a_l^{i}(\mathbf{x}) {\phi}_l^{i}(k_j),
    \label{eq:pca_local}
\end{equation}
where $\mu^{i}$, $a_l^{i}$ and $\phi_l^{i}$ are the mean, coefficient and PC at redshift $z_i$, respectively, and $n_\mathrm{PCA}^{i}$ is the number of PCs for $z_i$. The compressed output vector is then
\begin{equation}
    \begin{split}
        \mathbf{y}_\mathrm{c} = & \,[a_1^{(1)}, a_2^{(1)}, \ldots, a_{n_\mathrm{PCA}^{(1)}}^{(1)}, a_1^{(2)}, \ldots, a_{n_\mathrm{PCA}^{(2)}}^{(2)}, \ldots,\\
        &\, a_1^{(n_z)}, \ldots, a_{n_\mathrm{PCA}^{(n_z)}}^{(n_z)}\,],
    \end{split}
    \label{eq:pca_local_comp}
\end{equation}
with dimensionality $d_\mathrm{out}^\mathrm{loc} = \sum_{i=1}^{n_z} n_\mathrm{PCA}^{i}$.

Following Ref.~\cite{Knabenhans2021}, we determine the number of PCs based on the cumulative variance they explain. Specifically, we select the smallest value of $n_\mathrm{PCA}$ (or $n_\mathrm{PCA}^{i}$ for local PCA) such that the remaining unexplained variance is $<10^{-5}$. While we do not investigate how emulator performance varies with this threshold in the present work, we note that the optimal choice is likely data-dependent. As such, it is generally advisable to assess emulator accuracy across a range of variance thresholds prior to finalizing the compression scheme.

PCA is applied to the output for each NN prior to training, i.e., to both $NN_\mathrm{L}$ and $NN_\mathrm{LH}$ in the 2-step architecture (original and modified versions). We implement PCA using the \textsc{scikit-learn} library~\cite{scikit-learn}.

\subsubsection{\label{sec:training}Neural network training}

Here we describe how we train an NN, assuming the architecture of the NN is given (i.e., with the number of layers and layer widths are pre-defined).
Training an NN is essentially a process of minimizing the discrepancy between the predicted output and the true output, which is usually done by minimizing a loss function through iteratively updating the weights ($\mathbf{W}$) and biases ($\mathbf{b}$) of the NN. \textsc{PyTorch}~\cite{Paszke2019} is used to implement the NNs in this work.

Suppose we have a training data set $\mathcal{T} = \{(\mathbf{x}^{i}, \mathbf{y}^{i}): i=1,2,\ldots,N_\mathrm{train}\}$, where $N_\mathrm{train}$ is the number of training samples, $\mathbf{x}^{i}$ is the $i$th input, and $\mathbf{y}^{i}$ is the corresponding output, and a separate validation set $\mathcal{T}_\mathrm{val} = \{(\mathbf{x}_\mathrm{val}^{i}, \mathbf{y}_\mathrm{val}^{i}): i=1,2,\ldots,N_\mathrm{val}\}$, where $N_\mathrm{val}$ is the number of validation samples. A good model should not only fit the training data well but also generalize well to unseen data. To achieve this, a loss function that can prevent a model from being too complex is needed. In this work, we employ a regularized loss function of the form
\begin{equation}
        \mathcal{L} (\mathbf{W}, \mathbf{b}) =\mathcal{L}_\mathrm{train} (\mathbf{W}, \mathbf{b}) + \lambda  ||\mathbf{W}||_2^2,
    \label{eq:loss}
\end{equation}
where $\mathcal{L}_\mathrm{train}$ is the training loss, measuring the distance between the predicted output and the training output data, and the second term is the regularization term, which penalizes large weights to prevent overfitting. The regularization parameter $\lambda$ is a hyperparameter that controls the strength of the regularization. We use the mean squared error (MSE) as the training loss, i.e.,
\begin{equation}
    \mathcal{L}_\mathrm{train} (\mathbf{W}, \mathbf{b}) = \frac{1}{N_\mathrm{train}} \sum_{i=1}^{N_\mathrm{train}} ||f_\mathrm{NN}(\mathbf{x}^{i};\mathbf{W},\mathbf{b}) - \mathbf{y}^{i}||_2^2,
    \label{eq:loss_train}
\end{equation}
where $f_\mathrm{NN}(\mathbf{x}^{i};\mathbf{W},\mathbf{b})$ is the predicted output of the NN with weights $\mathbf{W}$ and biases $\mathbf{b}$ at $\mathbf{x}^{i}$. Note that the loss is computed using PCA coefficients instead of the raw output itself. The loss function is minimized using the \textsc{AdamW} optimizer~\cite{Loshchilov2017}.\footnote{
    The \textsc{AdamW} optimizer is a variant of the \textsc{Adam} optimizer~\cite{Kingma2014} that decouples weight decay from the optimization process. However, we use explicit L2 regularization instead of weight decay in this work, so the \textsc{AdamW} optimizer behaves like the Adam optimizer.} The activation function used in the hidden layers is the SiLU function, which is a special case of the Swish function~\cite{Ramachandran2017} with $\beta=1$:
\begin{equation}
    \mathrm{SiLU}(x) = x \cdot \sigma(x) = \frac{x}{1 + e^{-x}},
\end{equation}
where $\sigma(x)$ refers to the sigmoid function.

We define the validation loss, $\mathcal{L}_\mathrm{val}$, in a similar way as the training loss, but replacing the training data with the validation data in Eq.~(\ref{eq:loss_train}).

A dynamically decreasing learning rate (LR) schedule is implemented to stabilize the training process. An initial LR is set and decreased if $\mathcal{L}+\mathcal{L}_\mathrm{val}$ does not improve for a certain number of epochs (patience). The schedule parameters, including the initial LR and patience, can be adjusted for different training runs.
% Note that the stopping criterion can optionally be based on $\mathcal{L}$ solely, since we do not rely on the validation loss but the regularization term to prevent overfitting for a given set of hyperparameters. The validation loss, however, will be used to determine the best hyperparameters in the hyperparameter optimization process (see Sec.~\ref{sec:hyper_opt}).

As a function of a large number of variables (the weights and biases), $\mathcal{L}$ can be very complex and have many local minima. The optimizer may converge to suboptimal solutions (bad local minima). To mitigate this, we perform multiple training runs with different random seeds for initialization, and the best model with the lowest loss is retained.

\subsubsection{\label{sec:hyper_opt}Hyperparameter optimization}

Our objective is to identify the optimal set of hyperparameters for the NN that minimizes the combined training and validation losses, thereby balancing underfitting and overfitting. The hyperparameters subject to optimization include the number of hidden layers $L$, the number of neurons per layer $M$ (assumed uniform across layers), and the regularization parameter $\lambda$.

We perform Bayesian optimization implemented with \textsc{Hyperopt}~\cite{Bergstra2013} in two stages. In the first stage, we perform a coarse search over a large space of hyperparameters, and in the second stage, we perform a refined search over a smaller region. The initial hyperparameter ranges used in this work\footnote{We chose the prior range empirically, picking values so that the optimal hyperparameters were not at the boundaries. The ranges can be data dependent and may be adjusted for specific problems.} are given in Table~\ref{tab:hyper_opt}.
For the first stage, we use a uniform prior for $L$ and $M$, and a log-uniform prior for $\lambda$. The ranges of the hyperparameters are chosen to be wide enough to cover a large space of hyperparameters. In the second stage, $L$ is fixed to the best value found in the first stage, since a different $L$ will lead to a significantly different NN that is unlikely to result in a better performance.\footnote{We empirically confirm that none of the best hyperparameter sets found in the first stage led to a better performance when $L$ was changed in our tests.} The prior for $M$ follows $\mathcal{U}(\left\{M_1 -16+2q : q=0,1,\ldots,16\right\})$, where $M_1$ is the best value found in the first stage. This defines a uniform prior over 25 integers centered at $M_1$ with a step size of 2. The prior for $\lambda$ is defined as $\mathcal{LU}(\lambda_1/2,2\lambda_1)$, where $\lambda_1$ is the best value found in the first stage.

\begin{table}%[b]%The best place to locate the table environment is directly after its first reference in text
    \caption{\label{tab:hyper_opt}%
    Ranges of hyperparameters used in the first stage of the hyperparameter optimization process. The prior for $M$ is uniform over integers from 16 to 512 in steps of 16.\footnote{$\mathcal{U}(\{\})$ denotes the discrete uniform distribution, and $\mathcal{LU}$ denotes the log-uniform distribution.}
}
    \begin{ruledtabular}
    \begin{tabular}{lcc}
    {Hyperparameter}&
    {Prior}\\
    \colrule
    $L$ & $\mathcal{U}(\{1,2,3,4,5,6,7\})$\\
    $M$ & $\mathcal{U}(\{16,32,48,\ldots,512\})$\\
    $\lambda$ & $\mathcal{LU}(10^{-9}, 5\times10^{-6})$\\
    \end{tabular}
    \end{ruledtabular}
\end{table}

Evaluating a point in the hyperparameter space involves training and validating the NN with the given hyperparameters. Notice that {\scriptsize Goku-W} does not have a separate validation set of HF data, so we will use leave-one-out cross-validation (LOOCV) to evaluate the performance of the emulator. LOOCV is a special case of $k$-fold cross-validation~\cite{Kohavi1995} with $k = N_\mathrm{train}$. In the next section, we detail how a given set of hyperparameters is evaluated with $k$-fold training and validation, for $NN_\mathrm{L}$ and $NN_\mathrm{LH}$, respectively. We have confirmed that the LOOCV performance is consistent with the performance on a separate test set in Appendix~\ref{app:separate_test}, where we trained an emulator based on the {\scriptsize Goku-pre-N} simulations~\cite{Yang2025} and tested it on the available test set.

\subsubsection{\label{sec:k_fold}k-fold training and validation}

$k$-fold cross-validation is a technique to estimate the performance of a model by splitting the training data into $k$ subsets (folds). The model is trained on $k-1$ folds and validated on the remaining fold. This process is repeated $k$ times, with each fold being used as the validation set once. The final performance is obtained by averaging the performance over all $k$ folds.
The quantity we minimize in the hyperparameter optimization process (Sec.~\ref{sec:hyper_opt}) is the mean of the training and validation losses, i.e., a combined loss as a function of the hyperparameters:
\begin{equation}
    \Phi(L,M,\lambda)= \frac{1}{2k}\sum_{i=1}^{k} \left[\Phi_{\mathrm{train},i}(L,M,\lambda) + \Phi_{\mathrm{val},i}(L,M,\lambda)\right],
    \label{eq:loss_comb}
\end{equation}
where $\Phi_{\mathrm{train}}(L,M,\lambda) = \min_{\mathbf{W},\mathbf{b}} \mathcal{L}_\mathrm{train}(L,M,\lambda;\mathbf{W},\mathbf{b})$ (the minimum training loss from Eq.~(\ref{eq:loss_train})) and $i$ is the index of the fold. $\Phi_{\mathrm{val},i}(L,M,\lambda)$ is defined in a similar way as $\Phi_{\mathrm{train},i}(L,M,\lambda)$ but with the training loss replaced by the validation loss.

For $NN_\mathrm{LH}$, the data set has $n_\mathrm{H}$ samples, and we split the data into $k = n_\mathrm{H}$ folds. In each iteration, we use $n_\mathrm{H}-1$ samples for training and 1 sample for validation. In addition, $n_\mathrm{seed}^\mathrm{LH} = 5$ random seeds are used to initialize the weights and biases of the NN for each fold training to avoid bad local minima.

\begin{figure}
    % \centering
    \includegraphics[width=.85\linewidth,trim=350 65 1185 370,clip]{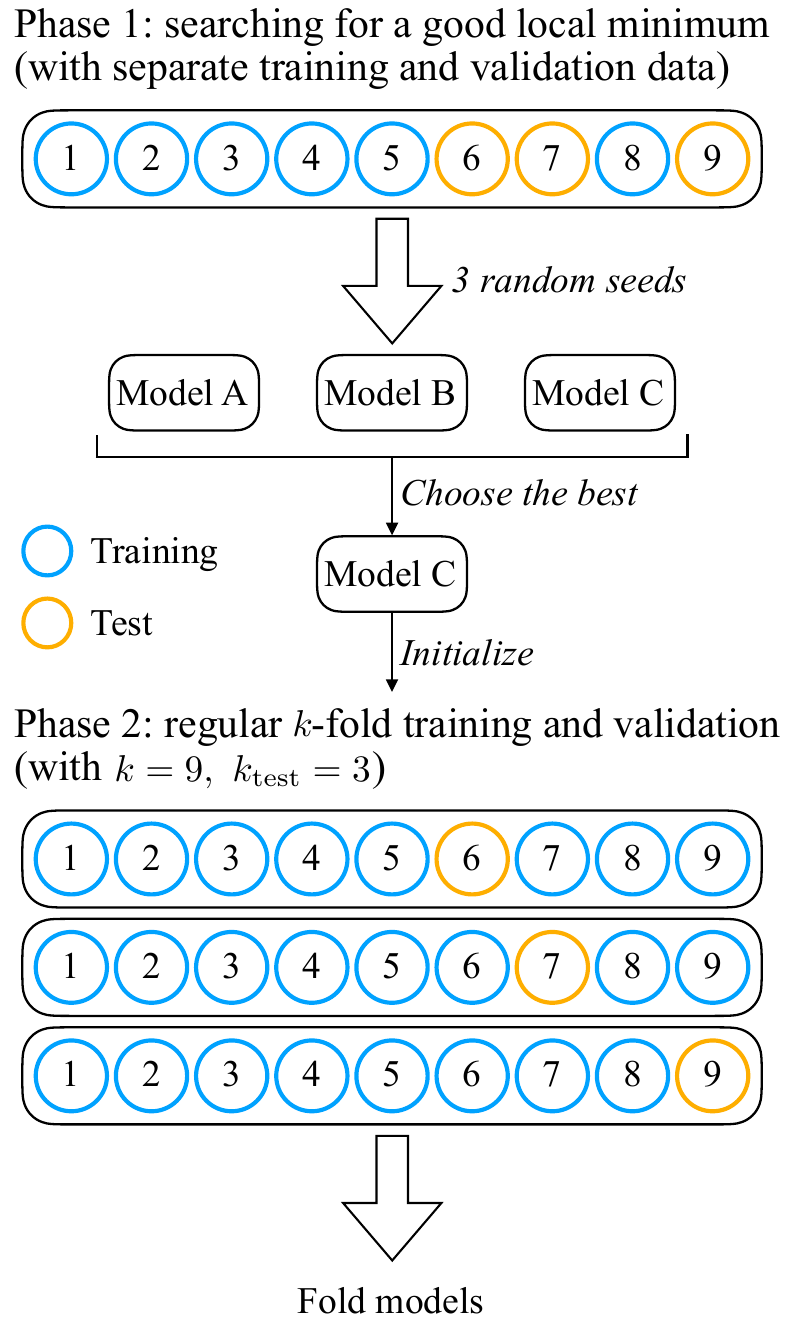} 
    \caption{\label{fig:2-phase}Illustration of the two-phase $k$-fold training and cross-validation strategy for $NN_\mathrm{L}$, assuming a total of 9 samples, of which 3 (orange circles) are supposed to be tested against (i.e., the HF cosmologies). In phase 1, the model is trained on the remaining 6 samples (blue circles) using 3 separate runs with different random seeds, and validated on the 3 held-out test samples. In phase 2, we perform regular $k$-fold training and validation, with the initial model (weights and biases) set to the best model found in phase 1.
    }
\end{figure}

Likewise, for $NN_\mathrm{L}$, the data set has $n_\mathrm{L}$ samples, and we split the data into $k = n_\mathrm{L}$ folds. However, only the $n_\mathrm{H}$ HF cosmologies should be tested on for our purpose, i.e., we only need to iterate over the $n_\mathrm{H}$ folds that leave the HF cosmologies out in training. We can take advantage of this feature and use a 2-phase training strategy, where a good local minimum is found in the first phase and then used as the common initial model for the second phase of training for each fold. This is much more efficient than the regular training technique we use for training $NN_\text{LH}$, since there will be no need to search for good minima for every fold by trying different random seeds independently. Specifically, in the first phase, the LF cosmology-only data (samples with HF cosmologies are excluded), $\mathcal{T}^\mathrm{L}_{1,\text{train}} = \left\{(\mathbf{x}^{i},\mathbf{y}^{i}) : 1\le i \le n_\mathrm{L}, \mathbf{x}^{i} \notin \mathcal{X}^\mathrm{H}  \right\}$, is used as the training set, and the LF data with the HF cosmologies, $\mathcal{T}^\mathrm{L}_{1,\text{val}} = \mathcal{T}^\mathrm{L} \setminus \mathcal{T}^\mathrm{L}_{1,\text{train}}$, is used as the validation set. We train the NN with $n_\mathrm{seed}^\mathrm{L} = 15$ random seeds for initialization in the first phase. The best model found in the first phase is then used to set the initial state of the NN for the second phase of training. We illustrate the process of the 2-phase training using a simple example in Fig.~\ref{fig:2-phase}. A bonus of this approach is that the second phase will only take a small number of epochs to converge, and is thus quite efficient in computation.\footnote{In practice, we also set the initial learning rate in the second phase to be equal to the final learning rate from the first phase to avoid jumping to other local minima.}
The second phase is the target $k$-fold training and validation, i.e., for each fold, the training set is $\mathcal{T}^{\mathrm{L},j}_{2,\text{train}} = \left\{(\mathbf{x}^{i},\mathbf{y}^{i}) : 1\le i \le n_\mathrm{L}, i\neq j, \mathbf{x}^{j} \in \mathcal{X}^\mathrm{H}  \right\}$, where $j$ is the index of the fold (also the HF cosmology), and the validation set is $\mathcal{T}^{\mathrm{L},j}_{2,\text{val}} = \left\{(\mathbf{x}^{(j)},\mathbf{y}^{(j)})\right\}$ (i.e., the point left out).

For the best-performing set of hyperparameters, we initialize the final model training using the fold model with the median regularized loss. To prevent overfitting during this final training step, we impose a lower bound on the training loss: specifically, the final training loss is not allowed to fall below 80\% of the median training loss observed across the folds. This threshold has proven effective in practice, as we have verified that the final model's performance remains consistent with the LOOCV results (see Appendix~\ref{app:separate_test}). Nevertheless, a more comprehensive evaluation of this thresholding strategy could be pursued in future work.

The 2-phase training strategy ensures that all the fold models fall into the same local minimum, and the validation error should be a better representative of the generalization error for the final model (also in the same local minimum) trained on the full LF data set compared to regular $k$-fold validation. Note that the HF cosmologies must be excluded from the training set in the first phase, though that phase is just for the sake of local minima searching instead of final validation. Because otherwise, the model for initialization would have memorized the data we are supposed to test on, and validation in the second phase would be invalid. This is also the reason why we cannot use the 2-phase strategy for $NN_\mathrm{LH}$ (no data available other than the test points) but have to try multiple random seeds for each fold training. 

% \subsubsection{\label{sec:add_opt}Additional optimization}

% Once the best hyperparameters are found, we perform additional optimization of the NN by training the NN with more random seeds for initialization, which may help to find a better local minimum. We use 20 times more random seeds in this phase than in the hyperparameter optimization process, i.e., for $NN_\mathrm{L}$, we use $n_\mathrm{seed}^\mathrm{L,add} = 300$ random seeds for initialization in the first round, and for $NN_\mathrm{LH}$, we set $n_\mathrm{seed}^\mathrm{LH,add} = 100$.

\subsection{\label{sec:compare}Comparative Study Design}

The techniques evaluated in this work are summarized in Table~\ref{tab:compare}. To assess the effectiveness of each technique, we design a comparative study with a series of different approaches for emulator construction. These approaches are distinct combinations of the techniques we mentioned above. The configurations for each approach are defined in Table~\ref{tab:approaches}. 

\begin{table*}
    \caption{\label{tab:compare}%
    Techniques considered in this work. The numbers 0, 1, and 0+ (if applicable) refer to the choices of the strategies, e.g., choice 0 represents the original 2-step model for the MF NN architecture. $n_\text{trial}$ is the number of trials in the coarse search stage of the hyperparameter optimization process, and $n_\text{trial}^\text{tune}$ is the number of trials in the fine-tuning stage. ``1-stage'' means no fine-tuning. For the training of $NN_\mathrm{L}$, ``1-phase'' refers to regular $k$-fold training and validation ($n_\text{seed}=1$ by default).}
    \begin{ruledtabular}
    \begin{tabular}{ccccc}
    Choice & {MF NN architecture} & PCA & Hyperparameter optimization & Training of $NN_\text{L}$\\
    \colrule
    0 & Original 2-step & Global (all-$z$) & 1-stage with $n_\text{trial}=80$ & 1-phase with $n_\text{seed}=1$\\
    1 & Modified 2-step & Local (per-$z$) & 2-stage with $n_\text{trial}=80$ and $n_\text{trial}^\text{tune}=40$ & 2-phase with $n_\text{seed}=15$\\
    0+ & & & & 1-phase with $n_\text{seed}=3$\\

    \end{tabular}
    \end{ruledtabular}
\end{table*}

\begin{table}
    \caption{\label{tab:approaches}%
    Approaches tested in this work. MFA, PCA, HO, and $NN_\mathrm{L}$ are short for the column names in Table~\ref{tab:compare}. The numbers 0, 1 and 0+ refer to the techniques described in Table~\ref{tab:compare} for each column. For PCA, we also list the number of PCs used for $NN_\mathrm{L}$ and $NN_\mathrm{LH}$, respectively, in parentheses.}
    \begin{ruledtabular}
    \begin{tabular}{ccccc}
    Approach & {MFA} & PCA ($n_\mathrm{PCA}^\mathrm{L}, n_\mathrm{PCA}^\mathrm{LH}$) & HO & $NN_\text{L}$\\
    \colrule
    \texttt{Base} & 0 & 0 (21, 12) & 0 & 0\\
    \texttt{Arch-0} & 0 & 1 (50, 42) & 0 & 0\\
    \texttt{PCA-0} & 1 & 0 (21, 20) & 0 & 0\\
    \texttt{Mid} & 1 & 1 (50, 118) & 0 & 0\\
    \texttt{NNL-1} & 1 & 1 (50, 118) & 0 & 1\\
    \texttt{NNL-0+} & 1 & 1 (50, 118) & 0 & 0+\\
    \texttt{Optimal} & 1 & 1 (50, 118) & 1 & 1\\

    \end{tabular}
    \end{ruledtabular}
\end{table}

\texttt{Mid} serves as the reference approach, which uses the modified 2-step architecture, separate PCA for each redshift, but does not include hyperparameter fine-tuning and 2-phase training of $NN_\mathrm{L}$. \texttt{Base} is the most basic approach, with the original 2-step architecture, global PCA, and no additional optimization strategies. The most advanced approach, \texttt{Optimal}, incorporates all enhanced techniques. The remaining approaches differ from \texttt{Mid} by altering only one component, allowing us to isolate the contribution of each technique. For example, \texttt{Arch-0} uses the original 2-step architecture but keeps the other techniques the same as \texttt{Mid}. \texttt{HO-2} uses 2-stage hyperparameter optimization without changing other components.

By comparing \texttt{HO-2} and \texttt{Mid}, we will see the effect of hyperparameter fine-tuning. However, it is not a strictly fair comparison, since they would have significant differences in compute time. To make a fair comparison, we define \texttt{HO-3}, which uses the same 1-stage hyperparameter optimization as \texttt{Mid} but with a larger number of trials, $n_\mathrm{trial}=120$, ensuring that the total compute time is similar to \texttt{HO-2}. Similarly, while comparing \texttt{NNL-1} and \texttt{Mid} will show the effect of 2-phase training of $NN_\mathrm{L}$, we also define \texttt{NNL-0+} which uses the same 1-phase training as \texttt{Mid} (which does not try multiple seeds) but with a larger number of random seeds, $n_\mathrm{seed}=3$, leading to a similar compute time as \texttt{NNL-1}. Although we do not present a detailed quantitative comparison of compute times across all approaches, we note that training each emulator requires less than 24 hours on a single Grace-Hopper node of the Vista supercomputer\footnote{\url{https://tacc.utexas.edu/systems/vista/}}. This cost is negligible relative to the computational expense of running the simulations themselves in the context of simulation-based emulation.

The results of the comparative study will be shown in Sec.~\ref{sec:results}, where we will compare the performance of the emulators built with different approaches and also discuss the impact of each technique at the level of the component NNs ($NN_\mathrm{L}$ and $NN_\mathrm{LH}$).

\section{\label{sec:results}Results}

We present the results of the comparative study in this section. The models trained with different approaches are evaluated using LOOCV, with the validation error defined as the relative mean absolute error (rMAE) of the predicted power spectrum compared to the true power spectrum, denoted as $\Phi_\text{rMAE}$. For clarity, each model is identified by the name of the approach used in its construction (e.g., the model trained with the \texttt{Base} approach is referred to as \texttt{Base}).

\begin{figure*}
    \centering
    \includegraphics[width=\textwidth]{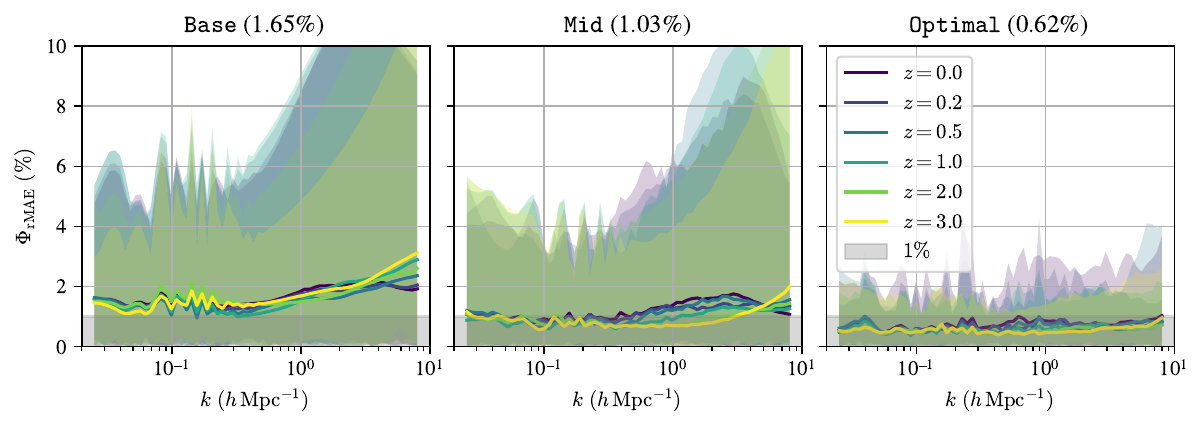}
    \caption{LOO errors of the emulators built with approaches \texttt{Base}, \texttt{Mid}, and \texttt{Optimal}. Redshifts are color coded. The solid lines are the error averaged over cosmologies, and the corresponding shaded regions indicate the range of individual cosmologies. The gray-shaded area marks the region where the error is less than 1\%. Each model is titled with the name of the approach and its overall validation error.}
    \label{fig:loo_3cases}
\end{figure*}

\begin{figure}
    \centering
    \includegraphics[width=\linewidth]{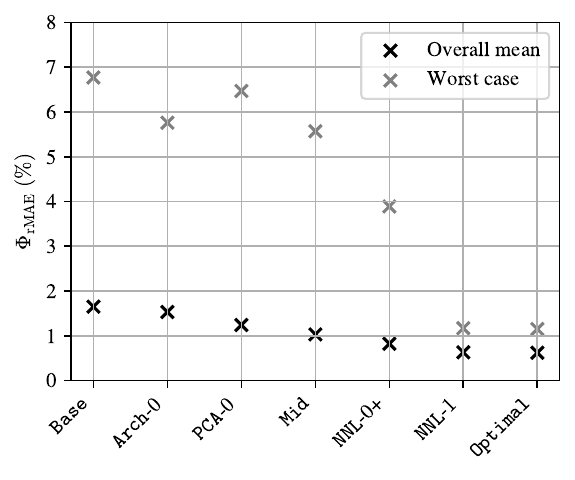}
    \caption{Summary comparison of the LOO errors of the emulators built with different approaches. Black crosses indicate the mean validation error of each approach, while gray crosses show the worst-case errors (the maximal error over all test points).}
    \label{fig:all-in-one}
\end{figure}

Figure~\ref{fig:loo_3cases} shows the validation errors as functions of $k$ and $z$ for \texttt{Base}, \texttt{Mid}, and \texttt{Optimal}. We found that even the basic model, \texttt{Base}, achieves a validation error significantly lower than {\scriptsize GokuEmu}'s 3\% error (see Fig.~13 of Ref.~\cite{Yang2025}). This suggests that NNs may be better-suited for emulation tasks involving large training sets and high-dimensional parameter spaces than GPs. The improvement may be accounted for by more efficient training that allows more intensive hyperparameter optimization, though PCA could have also contributed to the performance improvement.
Compared to \texttt{Base}, \texttt{Mid} achieves a significant improvement in accuracy, with an overall validation error of 1.03\% (compared to 1.65\% for \texttt{Base}), attributed to the modified 2-step architecture and the local PCA strategy.\footnote{Similar compute times were used to train these two models.} The improvement is observed across all redshifts and wavenumbers, though the worst-case error is still much higher than the average. The validation error of \texttt{Optimal} is less than 1\% for all redshifts and almost all wavenumbers, with an overall mean of 0.62\%, which is a further improvement over \texttt{Mid} resulting from the changes in hyperparameter optimization and training of $NN_\mathrm{L}$. Not only is the overall validation error reduced, but the worst-case error is also considerably lower than that of \texttt{Mid} (a reduction by a factor of 5 will be seen in Fig.~\ref{fig:all-in-one}).

A summary comparison of the LOO errors of the emulators built with different approaches is shown in Fig.~\ref{fig:all-in-one}, with both the overall mean error and the worst-case error shown. The error of \texttt{Mid} is lower than that of \texttt{Arch-0} and \texttt{PCA-0}, indicating both the modified 2-step architecture and the local PCA strategy are effective in improving the performance of the emulator, while the 2-step architecture leads to a larger improvement than the local PCA data compression strategy. While the aforementioned techniques improve the overall mean error, the worst-case error is still high.\footnote{We remind the reader that the fold models are trained independently under the standard LOO scheme (i.e., 1-phase training) and they may converge to different local minima. The worst case corresponds to a fold model which converges to a poor minimum.} A substantial reduction in the worst-case error is seen when implementing a 2-phase training strategy, \texttt{NNL-1}, which allows a large number of local minima to be explored efficiently. \texttt{NNL-0+} also shows an improvement over \texttt{Mid}, but the improvement is not as significant as that of \texttt{NNL-1}, despite consuming a similar amount of compute time (essentially because \texttt{NNL-0+} tried less random seeds than \texttt{NNL-1}). We emphasize that the differences between \texttt{Mid}, \texttt{NNL-0+} and \texttt{NNL-1} arise primarily from the different numbers of random seeds used during training. The 2-phase strategy (\texttt{NNL-1}) is designed to improve training efficiency (see also Sec.~\ref{sec:k_fold}), rather than to push the intrinsic accuracy limit of the model, i.e., regular training and cross-validation with $n_\mathrm{seed}=15$ would yield a similar error level to \texttt{NNL-1}, but at a significantly higher computational computational cost.\footnote{We note that, in principle, the 2-phase method could also be applied to the GPs in Ref.~\cite{Yang2025} to shorten training (i.e., hyperparameter optimization in this case) times for LOOCV.} \texttt{Optimal} achieves slightly lower errors than \texttt{NNL-1}, attributed to hyperparameter fine-tuning.\footnote{We have checked that simply increasing the number of trials (i.e., $n_\text{trial} > 80$) did not improve the performance, suggesting that it is the hyperparameter fine-tuning which is responsible for the (small) improvement in performance.}

In the following subsections, we present the effects of each technique in more detail, by comparing the performance of the component NNs ($NN_\mathrm{L}$ and $NN_\mathrm{LH}$) trained with different approaches. Figures~\ref{fig:arch_comp}, \ref{fig:pca_comp} and \ref{fig:nnl_comp} show the rMAE of the component NNs, defined as the LOO error of the predicted power spectrum compared to the true power spectrum. This ensures that $NN_\text{L}$ and $NN_\text{LH}$ are evaluated separately and independently. Specifically, in Figure~\ref{fig:arch_comp} and the lower panel of Figure~\ref{fig:pca_comp}, the component shown is $NN_\text{LH}$. Thus the input is the test cosmology and the true LF power spectrum, instead of the LF power spectrum predicted by $NN_\text{L}$. In the upper panel of Figure \ref{fig:pca_comp} and Figure \ref{fig:nnl_comp}, the component shown is $NN_\text{L}$. In this case, both the predicted power spectrum and the true power spectrum that it is tested against are LF power spectra.

\subsection{\label{sec:arch_result}Architecture: 2-Step vs. Modified 2-Step}

\begin{figure}
    \centering
    \includegraphics[width=\linewidth]{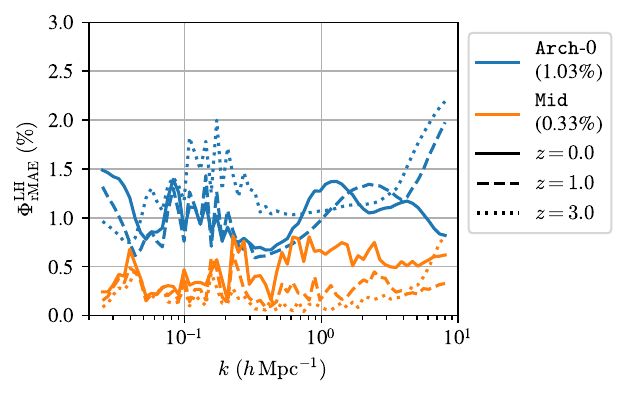}
    \caption{Comparison of the LF-to-HF correction NNs of the original (\texttt{Arch-0}) and the modified (\texttt{Mid}) 2-step architectures. Blue lines are LOO errors of \texttt{Arch-0}'s $NN_\text{LH}$, while orange lines are \texttt{Mid}'s. The solid, dashed, and dotted lines correspond to $z=0$, $1$, and $3$, respectively. The overall mean errors averaged over 6 redshifts are shown in the legends.}
    \label{fig:arch_comp}
\end{figure}

Figure~\ref{fig:arch_comp} compares the LF-to-HF correction NNs of the original and modified 2-step architectures. The validation error of \texttt{Mid} is shown to be significantly lower than that of \texttt{Arch-0} across redshifts and scales, with the average error reduced by a factor of $\sim 3$. In addition, \texttt{Mid}'s error decreases with increasing redshift (especially at small scales), which is consistent with our expectation that it is easier to learn the LF-to-HF correction at higher redshifts, where the spectrum is more linear and less affected by nonlinear effects. In contrast, \texttt{Arch-0} has a moderately larger error at $z=3$ than at $z=0$ and $1$. This suggests that the original architecture struggles to learn the correlation between the LF and HF power spectra and the information of the data is not fully exploited. The improvement is likely due to the aforementioned significantly reduced complexity of the NN (Sec.~\ref{sec:arch}) relative to the original architecture~\cite{Guo2022}.
 
\subsection{\label{sec:compress_result}Data Compression: Global vs. Local (PCA)}

\begin{figure}
    \centering
    \includegraphics[width=\linewidth]{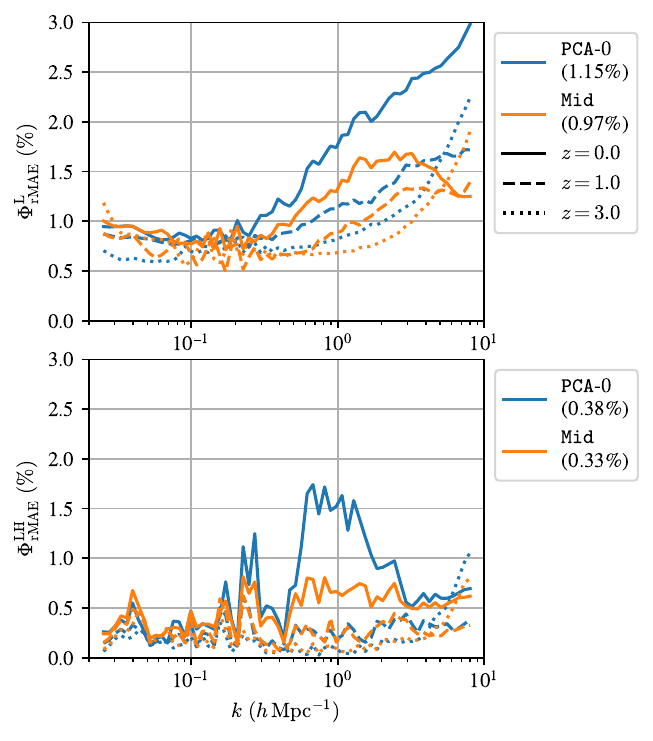}
    \caption{Comparison of the two data compression strategies: global PCA (\texttt{PCA-0}, in blue) and separate PCA for each redshift (\texttt{Mid}, in orange). The top and bottom panels show the LOO errors for $NN_\mathrm{L}$ and $NN_\mathrm{LH}$, respectively. The solid, dashed, and dotted lines correspond to $z=0$, $1$, and $3$, respectively. The overall mean errors averaged over 6 redshifts are shown in the legends.}
    \label{fig:pca_comp}
\end{figure}

From Fig.~\ref{fig:pca_comp}, we observe that the local PCA strategy (\texttt{Mid}) outperforms the global PCA strategy (\texttt{PCA-0}) for both $NN_\mathrm{L}$ and $NN_\mathrm{LH}$, which is likely because the global PCA is not as flexible as the local PCA in capturing redshift-dependent features of the spectrum.\footnote{We have confirmed that setting $n_\text{PCA}^\text{L}$ to 50 (the same as \texttt{Mid}) for \texttt{PCA-0} does not improve accuracy. So this improvement cannot be simply attributed to the increased number of PCs.} In particular, the improvement is more pronounced at $z = 0$ in both NNs, where the spectrum is more nonlinear. In Appendix~\ref{app:varying_nz}, we further verify that the local PCA strategy consistently outperforms the global PCA strategy when varying the number of redshift bins used in the data set.

We also note that $\Phi_\text{rMAE}^\text{L}$ is larger than $\Phi_\text{rMAE}^\text{LH}$ in both cases, indicating the uncertainty of the interpolation of the LF power spectrum in the parameter space dominates the overall error of the emulator, consistent with the findings of Ref.~\cite{Yang2025}. 

\subsection{\label{sec:LF_training}Training of the LF NN: Regular vs. 2-Phase}

\begin{figure}
    \centering
    \includegraphics[width=\linewidth]{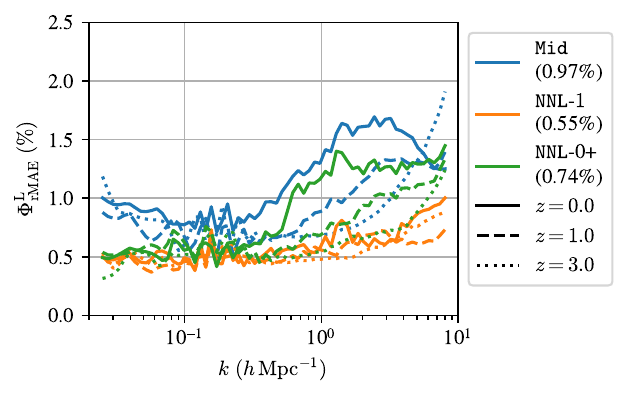}
    \caption{Comparison of the training strategies for $NN_\text{L}$. \texttt{Mid} (blue) and \texttt{NNL-0+} (green) use regular training, while \texttt{NNL-1} (orange) uses the 2-phase training strategy. \texttt{NNL-0+} tried more random seeds than \texttt{Mid} to match the compute time of \texttt{NNL-1}. Redshifts $z=0$, $1$, and $3$ are coded with solid, dashed, and dotted lines, respectively. The overall LOO errors averaged over 6 redshifts are shown in the legends.}
    \label{fig:nnl_comp}
\end{figure}

Fig.~\ref{fig:nnl_comp} compares the LF NNs trained with different strategies. Compared to \texttt{Mid}, \texttt{NNL-1} reduces the overall error significantly from 0.97\% to 0.55\%, improving performance across all redshifts and scales. When we simply increased the number of random seeds over \texttt{Mid} (\texttt{NNL-0+}), the worst-case error was about midway between \texttt{Mid} and \texttt{NNL-1}, despite a similar compute time. Regular training with more random seeds, e.g., 15 distinct seeds, might allow a performance similar to \texttt{NNL-1}, but it would take much longer to train the NN and the final model initialized by one of the fold models might not generalize as well as the model trained with the 2-phase strategy--the fold models might have fallen into different local minima, and the chosen model is not guaranteed to be the one with the best generalization performance.

\section{\label{sec:concl}conclusion}

We have developed  {\scriptsize T2N-MusE}, a multifidelity neural network framework for cosmological emulation, which is capable of building highly optimized regression models to predict summary statistics. This framework is characterized by a novel 2-step architecture, per-$z$ PCA for data compression, 2-stage hyperparameter optimization, and a 2-phase training strategy for the low-fidelity regression model. Compared with our earlier GP methods ~\cite{Yang2025}, this NN approach reduces the average validation error by a factor of more than 5 and the worst-case error\footnote{The worst-case error was not reported in Ref.~\cite{Yang2025}, but we have computed it here and find $8.81\%$ for {\scriptsize GokuEmu-W}.} by a factor of $\sim 8$  on the same data.\footnote{The training of {\scriptsize GokuEmu-W} used both the L1 and L2 nodes, while we only use L2 in this work. So a factor of 5 is a very conservative estimate.} 

In this work, we build emulators based on the nonlinear matter power spectra from the {\scriptsize Goku} simulations suite~\cite{Yang2025} using different combinations of the various techniques for a comparative study. The results show that all the techniques we proposed are effective in improving the performance of the emulator, although the effect of hyperparameter fine-tuning is modest.
The novel 2-step MF architecture reduces the complexity of the LF-to-HF correction NN, decreasing the error by a factor of $\sim 3$. The per-$z$ PCA strategy allows NNs to learn the redshift-dependent features of the statistics of interest more accurately, with accuracy improved by more than $10\%$ in both the LF NN and the LF-to-HF correction NN compared to the global PCA strategy.
The 2-stage hyperparameter optimization strategy moderately improves the performance of the emulator by fine-tuning the hyperparameters in a smaller space after a coarse search. The 2-phase training strategy for the LF NN efficiently finds a common local minimum for $k$-fold (or LOO) training and validation and substantially improves the worst-case error.

{\scriptsize T2N-MusE} realizes highly efficient training of NNs on large data sets with high-dimensional parameter spaces that traditional GP-based methods struggle with. This demonstrates the effectiveness of {\scriptsize T2N-MuSE} not only as a high-accuracy optimization scheme in its own right, but also as a general tool for upgrading existing emulators to higher performance or expanding their parameter space, all at significantly reduced computational costs.
We have rebuilt a production emulator for the matter power spectrum with {\scriptsize T2N-MusE} based on {\scriptsize Goku}, named {\scriptsize GokuNEmu}. {\scriptsize GokuNEmu} is the highest performing in existence, in terms of error, dimensionality, parameter coverage and inference speed, and is presented in Ref.~\cite{Yang2025b}. We will also apply this framework to build emulators for other summary statistics, such as the Lyman-$\alpha$ forest flux power spectrum~\cite{Bird2023} in future work. The code of {\scriptsize T2N-MusE} is publicly available at Ref.~\cite{T2N2025} for the community to use and extend.

\section*{Acknowledgments}
We thank Mark Achenbach for suggestions on our hyperparameter optimization code.
YY and SB acknowledge funding from NASA ATP 80NSSC22K1897. MFH is supported by the Leinweber Foundation and DOE grant DE-SC0019193. Computing resources were provided by Frontera LRAC AST21005.
The authors acknowledge the Frontera and Vista computing projects at the Texas Advanced Computing Center (TACC, \url{http://www.tacc.utexas.edu}) for providing HPC and storage resources that have contributed to the research results reported within this paper.
Frontera and Vista are made possible by National Science Foundation award OAC-1818253.

% Appendix section
\appendix

\section{\label{app:L1_optimal}Emulator trained on HF and L1}

\begin{figure}
    \centering
    \includegraphics[width=\linewidth]{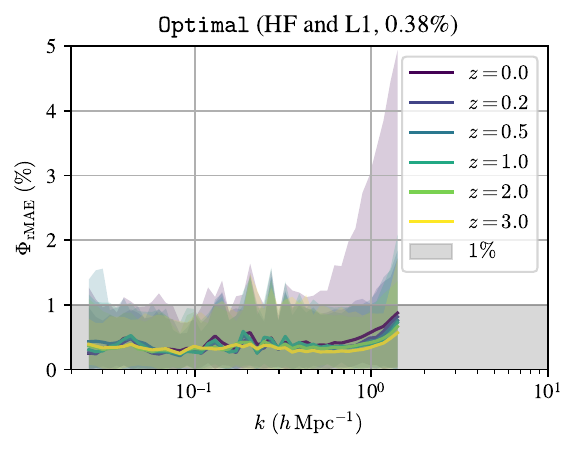}
    \caption{LOO error of an emulator trained on HF and L1 data with the \texttt{Optimal} approach. Solid lines are the error averaged over cosmologies, and the corresponding shaded regions indicate the range of individual cosmologies. The gray-shaded area marks the region where the error is less than 1\%. The title shows the overall mean error averaged over 6 redshift bins.}
    \label{fig:loocv_HFL1}
\end{figure}

We apply the \texttt{Optimal} approach to train an emulator on the HF and L1 nodes of {\scriptsize Goku-W} and show the LOOCV result in Fig.~\ref{fig:loocv_HFL1}. The data (matter power spectra) are truncated at $k=1.5h/\text{Mpc}$ here, since the L1 simulations do not include cosmological information at smaller scales (due to insufficient resolution) and including those scales would degrade the performance of the emulator. The mean error is less than 0.5\% for all redshifts across most scales, and the worst-case error is well controlled around 1\% except at the smallest scales where the L1 data are resolution limited. The performance is significantly better than {\scriptsize GokuEmu} (see Fig.~13 of Ref.~\cite{Yang2025}), which has a mean error between 1\% and 2\% and a worst-case error of $\gtrsim5\%$ over these scales and redshifts. This suggests that the NN approach proposed in this work is also effective at using the L1 data to improve emulator performance.
 
\section{\label{app:separate_test}LOOCV vs. Separate Test Set}

We train an emulator based on the preliminary simulation set, {\scriptsize Goku-pre-N}, using the \texttt{Optimal} approach and test it on the available test set. {\scriptsize Goku-pre-N} contains 297 pairs of LF simulations and 27 HF simulations in the training set and 12 HF simulations in the test set. Following the main text, we do not use L1 simulations in this study. The HF simulations evolve 300$^3$ particles in a box of size 100$\,\text{Mpc}/h$. For more details about the {\scriptsize Goku-pre-N} simulations, see Ref.~\cite{Yang2025}. 

\begin{figure}[ht]
    \centering
    \includegraphics[width=\linewidth]{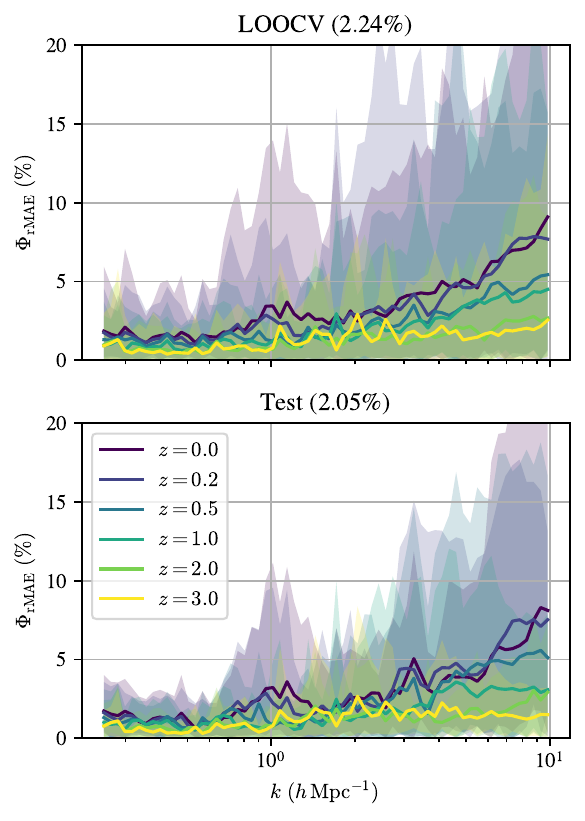}
    \caption{Comparison of LOO error (top) and test error (bottom) for the emulator trained on the {\scriptsize Goku-pre-N} simulations. The solid lines are the mean errors, and the shaded regions indicate the range of individual cosmologies. The redshifts are color coded, and the overall mean errors are shown in the titles of the panels.}
    \label{fig:loocv_vs_test_set}
\end{figure}

The LOO error and test error of the emulator are shown in the top and bottom panels of Fig.~\ref{fig:loocv_vs_test_set}, respectively. They are consistent with each other, with the test error being slightly lower than the LOO error. This indicates that the LOO cross-validation is a good representative of the generalization error of the final emulator trained on the full training set.

\section{\label{app:varying_nz}PCA: Global vs. Local with varying $n_z$}

\begin{figure}
    \centering
    \includegraphics[width=\linewidth]{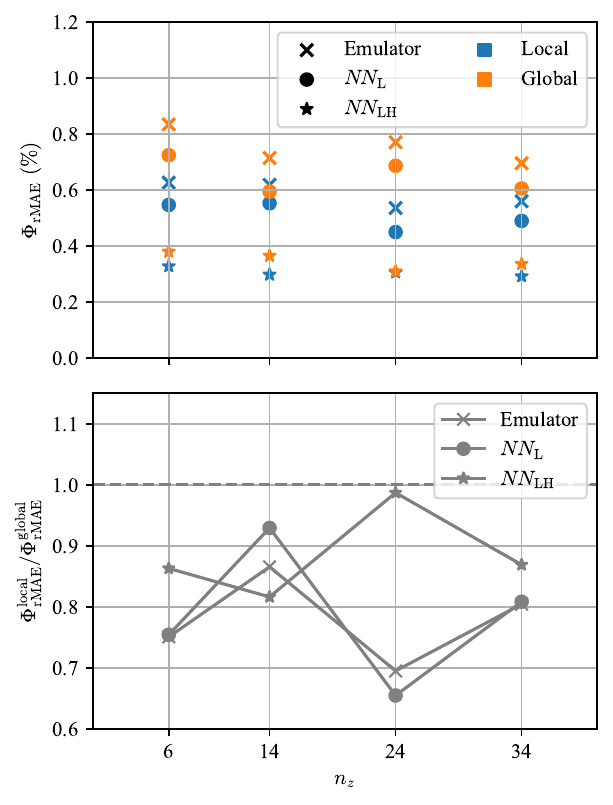}
    \caption{Comparison of global and local PCA strategies with varying number of redshift bins, $n_z$. Crosses, dots and stars show results for the emulators, $NN_\mathrm{L}$ and $NN_\mathrm{LH}$, respectively. Upper panel: LOO errors (mean) of the emulators and the component NNs trained with the local (blue) and global (orange) PCA strategies. Lower panel: Ratio of the errors of the local PCA strategy to those of the global PCA strategy. The dashed horizontal line indicates a ratio of 1.}
    \label{fig:local_vs_global_nz}
\end{figure} 

To further verify the advantage of the local PCA strategy over the global PCA strategy, we vary the number of redshift bins, $n_z$, used in the data set and compare the performance of the emulators and the component NNs trained with the two PCA strategies. Specifically, we consider $n_z = 6, 14, 24,$ and $34$, where the added redshifts (beyond the original 6 bins) are evenly spaced in $\ln a$ (with $a$ denoting the cosmic scale factor). For each case, we train the models using the modified 2-step MF architecture, the 2-phase $NN_\mathrm{L}$ training method, and the target PCA strategy (local or global), without hyperparameter fine-tuning (stage 2). In addition, we provide the number of PCs and hyperparameters used for each model in Table~\ref{tab:nz_pca} for readers who are interested.\footnote{Note that NN hyperparameters could change with optimization runs even with the same settings due to the stochastic nature of the optimization process. So we do not expect the hyperparameters have a clear trend with $n_z$.}

\begin{table*}
    \caption{Number of PCs and hyperparameters for each of the emulators trained with different PCA strategies and varying number of redshift bins, $n_z$. $L$ is the number of hidden layers, $M$ is the number of neurons per hidden layer, and $\lambda$ is the L2 regularization coefficient. The LF and LF-to-HF correction NNs are denoted by superscripts L and LH, respectively.}
    \begin{ruledtabular}
    \begin{tabular}{ccccccccc}
    $n_z$ & $n_\mathrm{PCA}^\mathrm{L}$ & $L^\mathrm{L}$ & $M^\mathrm{L}$ & $\lambda^\mathrm{L}$ & $n_\mathrm{PCA}^\mathrm{LH}$ & $L^\mathrm{LH}$ & $M^\mathrm{LH}$ & $\lambda^\mathrm{LH}$ \\
    \colrule
    Global: & & & & & & & & \\
    6 & 21 & 7 & 112 & $3.97\times10^{-9}$ & 20 & 7 & 208 & $8.84\times10^{-8}$ \\
    14 & 22 & 6 & 400 & $7.20\times10^{-7}$ & 20 & 5 & 16 & $2.86\times10^{-8}$ \\
    24 & 24 & 5 & 96 & $2.64\times10^{-7}$ & 20 & 6 & 16 & $1.14\times10^{-6}$ \\
    34 & 24 & 6 & 256 & $2.73\times10^{-9}$ & 20 & 6 & 16 & $1.57\times10^{-7}$ \\
    \colrule
    Local: & & & & & & & & \\
    6 & 50 & 7 & 272 & $1.91\times10^{-9}$ & 118 & 7 & 16 & $2.96\times10^{-8}$ \\
    14 & 111 & 5 & 336 & $1.40\times10^{-6}$ & 276 & 7 & 16 & $2.80\times10^{-8}$ \\
    24 & 189 & 6 & 240 & $1.93\times10^{-9}$ & 473 & 5 & 32 & $6.77\times10^{-9}$ \\
    34 & 266 & 6 & 256 & $2.03\times10^{-7}$ & 672 & 7 & 160 & $1.77\times10^{-7}$ \\

    \end{tabular}
    \end{ruledtabular}
    \label{tab:nz_pca}
\end{table*}

The results are shown in Fig.~\ref{fig:local_vs_global_nz}. We observe in the upper panel that the emulators/component NNs trained with the local PCA strategy consistently achieve lower LOO errors than those trained with the global PCA strategy across all tested values of $n_z$. The lower panel shows that the ratio of the errors of the local PCA strategy to those of the global PCA strategy remains below unity across all choices of $n_z$ and all models. For $n_z = 24$, the local PCA strategy yields a particularly notable reduction in emulator error, although its $NN_\mathrm{LH}$ component does not exhibit a comparable improvement. Overall, the local PCA strategy improves emulator accuracy by $\sim 10\%$--$30\%$ relative to the global strategy, with no clear monotonic dependence on $n_z$. These results further confirm the advantage of the local PCA approach in enhancing emulator performance.

\begin{figure}
    \centering
    \includegraphics[width=\linewidth]{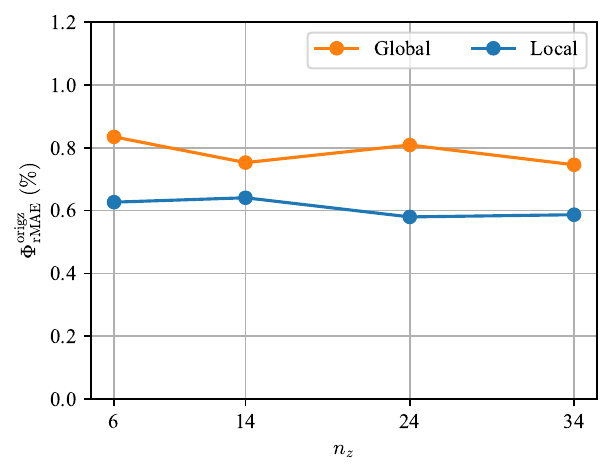}
    \caption{LOO error of the emulators evaluated at the original 6 redshifts as a function of $n_z$, the number of redshift bins used in the data set for training. Blue and orange lines represent results from the local and global PCA strategies, respectively.}
    \label{fig:local_global_nz_origz}
\end{figure} 

We note that the errors shown in the upper panel of Fig.~\ref{fig:local_vs_global_nz} are computed over all redshifts used in the data set for training. To compare the accuracy of the emulators trained with different $n_z$ values more rigorously, we evaluate their LOO errors at the original 6 redshifts only. The results are shown in Fig.~\ref{fig:local_global_nz_origz}. We observe that for both PCA strategies, the prediction error does not vary significantly with $n_z$, indicating that we can keep the same accuracy for each redshift even when increasing the number of redshift bins (which ensures smoother interpolations over redshift). In agreement with Fig.~\ref{fig:local_vs_global_nz}, the local PCA strategy consistently outperforms the global PCA strategy across all tested values of $n_z$.

\begin{figure*}
    \centering
    \includegraphics[width=\linewidth]{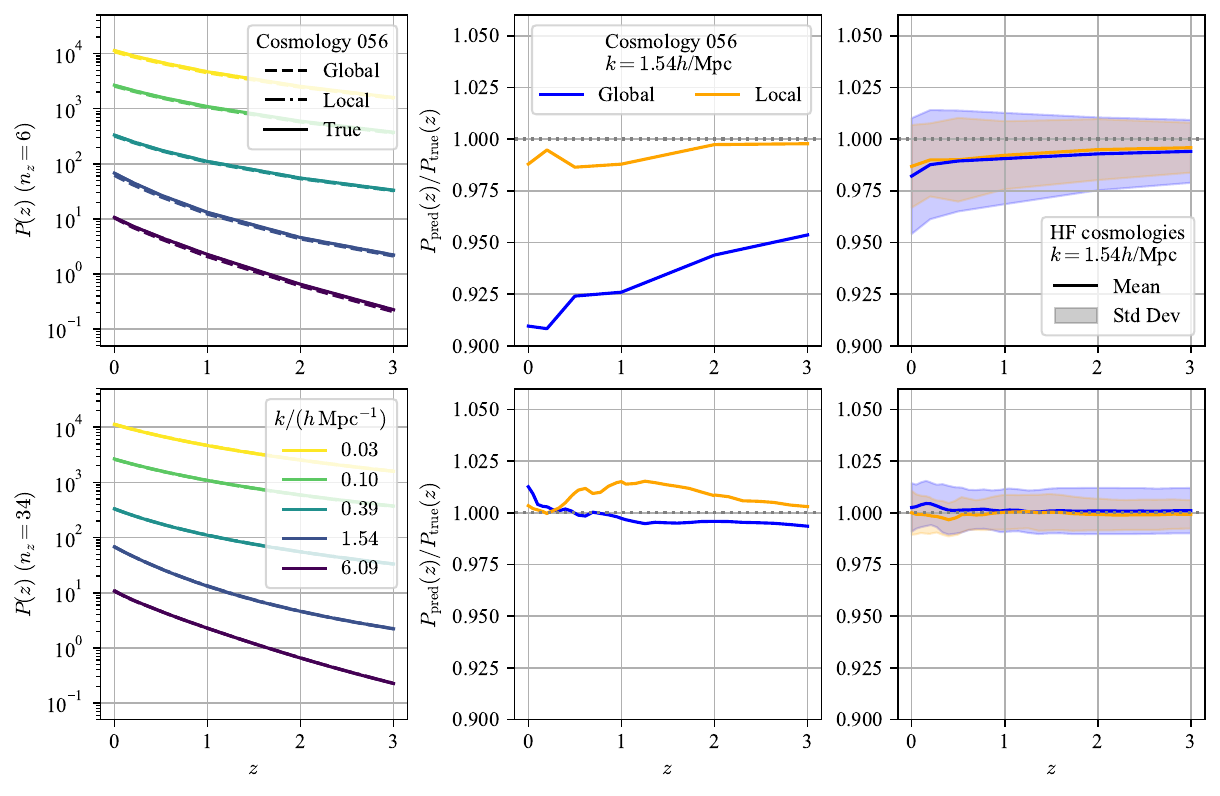}
    \caption{Comparison of the power spectrum as a function of redshift, $P(z)$, predicted by the emulators built with global and local PCA strategies against the true power spectrum. Left column: $P(z)$ of cosmology 56 (i.e., \texttt{Goku-W-0056}) at 5 different wavenumbers, $k=0.03, 0.10, 0.39, 1.54$ and $6.09h/\text{Mpc}$ (in different colors). Dashed, dash-dotted, and solid lines represent the predictions from the global PCA emulator, local PCA emulator, and the true power spectrum, respectively. Middle column: Ratio of the predicted $P(z)$ (at $k=1.54h\,\text{Mpc}^{-1}$) of cosmology 056 to the true $P(z)$ from the global (blue) and local (orange) PCA emulators. Right column: Mean ratio of the predicted $P(z)$ (at $k=1.54h\,\text{Mpc}^{-1}$) to the true $P(z)$ over all HF cosmologies from the global (blue) and local (orange) PCA emulators. Solid lines and shaded regions represent the mean and the standard deviation, respectively. The first and second rows present results for $n_z=6$ and $n_z=34$, respectively.}
    \label{fig:Pz_PCA_compare} 
\end{figure*}

We also examine the predicted power spectrum as a function of redshift, $P(z)$, from the emulators constructed using the two PCA strategies for different values of $n_z$. Figure~\ref{fig:Pz_PCA_compare} presents the results for $n_z=6$ (top row) and $n_z=34$ (bottom row). The left column shows $P(z)$ at five representative $k$ modes for a sample cosmology (056). The middle column displays the ratio of the predicted $P(z)$ at $k=1.54h/\text{Mpc}$, a scale where the local PCA strategy exhibits notably better performance than the global strategy (see Fig.~\ref{fig:pca_comp}), to the corresponding true $P(z)$ for cosmology 056, while the right column shows the mean ratio over all HF cosmologies. From the left column, we observe that the predictions are all smooth across redshift for both PCA strategies and both values of $n_z$. However, for $n_z=6$, the global PCA predictions systematically underestimate $P(z)$ at small scales (e.g., $k=1.54$ and $6.09h/\text{Mpc}$) regardless of redshift. This bias is clearly visible in the upper middle panel. In the middle column, we notice that the smoothness of the predictions from the local strategy is comparable to that from the global strategy, even though the local PCA is performed separately at each redshift without explicitly enforcing smoothness over redshift. The right column indicates that the prediction uncertainty from the local PCA strategy is better controlled than that from the global PCA strategy for both $n_z=6$ and $34$, consistent with the findings in Fig.~\ref{fig:local_vs_global_nz}. 

To summarize, both the global and local PCA strategies yield smooth predictions of $P(z)$ across redshifts. However, the local PCA strategy demonstrates superior accuracy compared to the global PCA strategy, irrespective of the number of redshift bins used.

\bibliography{MusE}% Produces the bibliography via BibTeX.

@ARTICLE{Guo2022,
       author = {{Guo}, Mengwu and {Manzoni}, Andrea and {Amendt}, Maurice and {Conti}, Paolo and {Hesthaven}, Jan S.},
        title = "{Multi-fidelity regression using artificial neural networks: Efficient approximation of parameter-dependent output quantities}",
      journal = {Computer Methods in Applied Mechanics and Engineering},
         year = 2022,
        month = feb,
       volume = {389},
        pages = {114378},
          doi = {10.1016/j.cma.2021.114378},
archivePrefix = {arXiv},
       eprint = {2102.13403},
 primaryClass = {math.NA},
       adsurl = {https://ui.adsabs.harvard.edu/abs/2022CMAME.389k4378G}
}

@Article{DESICollaboration2016,
       author = {{DESI Collaboration} and {Aghamousa}, Amir and {Aguilar et al.}, Jessica},
        title = "{The DESI Experiment Part I: Science,Targeting, and Survey Design}",
      journal = {arXiv e-prints},
         year = 2016,
        month = oct,
          eid = {arXiv:1611.00036},
        pages = {arXiv:1611.00036},
          doi = {10.48550/arXiv.1611.00036},
archivePrefix = {arXiv},
       eprint = {1611.00036},
 primaryClass = {astro-ph.IM},
       adsurl = {https://ui.adsabs.harvard.edu/abs/2016arXiv161100036D}
}

@Book{LSC2009,
  author        = {{Abell et al.}, Paul A.},
  title         = {LSST Science Book, Version 2.0},
  year          = {2009},
  month         = dec,
  archiveprefix = {arXiv},
  copyright     = {arXiv.org perpetual, non-exclusive license},
  doi           = {10.48550/ARXIV.0912.0201},
  eprint        = {0912.0201},
  primaryclass  = {astro-ph.IM},
  publisher     = {arXiv},
}

@Article{Laureijs2011,
  author        = {{Laureijs et al.}, R.},
  journal       = {arXiv e-prints},
  title         = "{Euclid Definition Study Report}",
  year          = {2011},
  month         = oct,
  pages         = {arXiv:1110.3193},
  archiveprefix = {arXiv},
  doi           = {10.48550/arXiv.1110.3193},
  eid           = {arXiv:1110.3193},
  eprint        = {1110.3193},
  primaryclass  = {astro-ph.CO},
  url           = {https://ui.adsabs.harvard.edu/abs/2011arXiv1110.3193L},
}

@Article{Akeson2019,
       author = {{Akeson et al.}, Rachel},
        title = "{The Wide Field Infrared Survey Telescope: 100 Hubbles for the 2020s}",
      journal = {arXiv e-prints},
         year = 2019,
        month = feb,
          eid = {arXiv:1902.05569},
        pages = {arXiv:1902.05569},
          doi = {10.48550/arXiv.1902.05569},
archivePrefix = {arXiv},
       eprint = {1902.05569},
 primaryClass = {astro-ph.IM},
       adsurl = {https://ui.adsabs.harvard.edu/abs/2019arXiv190205569A}
}

@Article{Heitmann2009,
  author    = {Heitmann, Katrin and Higdon, David and White, Martin and Habib, Salman and Williams, Brian J. and Lawrence, Earl and Wagner, Christian},
  journal   = {The Astrophysical Journal},
  title     = "{THE COYOTE UNIVERSE. II. COSMOLOGICAL MODELS AND PRECISION EMULATION OF THE NONLINEAR MATTER POWER SPECTRUM}",
  year      = {2009},
  issn      = {1538-4357},
  month     = oct,
  number    = {1},
  pages     = {156--174},
  volume    = {705},
  doi       = {10.1088/0004-637x/705/1/156},
  publisher = {American Astronomical Society},
}

@Article{Heitmann2010,
  author    = {Heitmann, Katrin and White, Martin and Wagner, Christian and Habib, Salman and Higdon, David},
  journal   = {The Astrophysical Journal},
  title     = "{THE COYOTE UNIVERSE. I. PRECISION DETERMINATION OF THE NONLINEAR MATTER POWER SPECTRUM}",
  year      = {2010},
  issn      = {1538-4357},
  month     = apr,
  number    = {1},
  pages     = {104--121},
  volume    = {715},
  doi       = {10.1088/0004-637x/715/1/104},
  publisher = {American Astronomical Society},
}

@Article{Heitmann2013,
  author    = {Heitmann, Katrin and Lawrence, Earl and Kwan, Juliana and Habib, Salman and Higdon, David},
  journal   = {The Astrophysical Journal},
  title     = "{THE COYOTE UNIVERSE EXTENDED: PRECISION EMULATION OF THE MATTER POWER SPECTRUM}",
  year      = {2013},
  issn      = {1538-4357},
  month     = dec,
  number    = {1},
  pages     = {111},
  volume    = {780},
  doi       = {10.1088/0004-637x/780/1/111},
  publisher = {American Astronomical Society},
}

@Article{Heitmann2016,
  author    = {Heitmann, Katrin and Bingham, Derek and Lawrence, Earl and Bergner, Steven and Habib, Salman and Higdon, David and Pope, Adrian and Biswas, Rahul and Finkel, Hal and Frontiere, Nicholas and Bhattacharya, Suman},
  journal   = {The Astrophysical Journal},
  title     = "{THE MIRA–TITAN UNIVERSE: PRECISION PREDICTIONS FOR DARK ENERGY SURVEYS}",
  year      = {2016},
  issn      = {1538-4357},
  month     = mar,
  number    = {2},
  pages     = {108},
  volume    = {820},
  doi       = {10.3847/0004-637x/820/2/108},
  publisher = {American Astronomical Society},
}

@Article{Lawrence2017,
  author    = {Lawrence, Earl and Heitmann, Katrin and Kwan, Juliana and Upadhye, Amol and Bingham, Derek and Habib, Salman and Higdon, David and Pope, Adrian and Finkel, Hal and Frontiere, Nicholas},
  journal   = {The Astrophysical Journal},
  title     = "{The Mira-Titan Universe. II. Matter Power Spectrum Emulation}",
  year      = {2017},
  issn      = {1538-4357},
  month     = sep,
  number    = {1},
  pages     = {50},
  volume    = {847},
  doi       = {10.3847/1538-4357/aa86a9},
  publisher = {American Astronomical Society},
}

@Article{DeRose2019,
  author    = {DeRose, Joseph and Wechsler, Risa H. and Tinker, Jeremy L. and Becker, Matthew R. and Mao, Yao-Yuan and McClintock, Thomas and McLaughlin, Sean and Rozo, Eduardo and Zhai, Zhongxu},
  journal   = {The Astrophysical Journal},
  title     = "{The Aemulus Project. I. Numerical Simulations for Precision Cosmology}",
  year      = {2019},
  issn      = {1538-4357},
  month     = apr,
  number    = {1},
  pages     = {69},
  volume    = {875},
  doi       = {10.3847/1538-4357/ab1085},
  publisher = {American Astronomical Society},
}

@Article{McClintock2019,
  author    = {McClintock, Thomas and Rozo, Eduardo and Becker, Matthew R. and DeRose, Joseph and Mao, Yao-Yuan and McLaughlin, Sean and Tinker, Jeremy L. and Wechsler, Risa H. and Zhai, Zhongxu},
  journal   = {The Astrophysical Journal},
  title     = "{The Aemulus Project. II. Emulating the Halo Mass Function}",
  year      = {2019},
  issn      = {1538-4357},
  month     = feb,
  number    = {1},
  pages     = {53},
  volume    = {872},
  doi       = {10.3847/1538-4357/aaf568},
  publisher = {American Astronomical Society},
}

@Article{Zhai2019,
  author    = {Zhai, Zhongxu and Tinker, Jeremy L. and Becker, Matthew R. and DeRose, Joseph and Mao, Yao-Yuan and McClintock, Thomas and McLaughlin, Sean and Rozo, Eduardo and Wechsler, Risa H.},
  journal   = {The Astrophysical Journal},
  title     = "{The Aemulus Project. III. Emulation of the Galaxy Correlation Function}",
  year      = {2019},
  issn      = {1538-4357},
  month     = mar,
  number    = {1},
  pages     = {95},
  volume    = {874},
  doi       = {10.3847/1538-4357/ab0d7b},
  publisher = {American Astronomical Society},
}

@Article{Smith2019,
  author    = {Smith, Robert E and Angulo, Raul E},
  journal   = {Monthly Notices of the Royal Astronomical Society},
  title     = "{Precision modelling of the matter power spectrum in a Planck-like Universe}",
  year      = {2019},
  issn      = {1365-2966},
  month     = apr,
  number    = {1},
  pages     = {1448--1479},
  volume    = {486},
  doi       = {10.1093/mnras/stz890},
  publisher = {Oxford University Press (OUP)},
}

@article{Knabenhans2019,
    author = {{Euclid Collaboration} and {Knabenhans}, Mischa and {Stadel et al.}, Joachim},
    title = "{Euclid preparation: II. The EuclidEmulator – a tool to compute the cosmology dependence of the nonlinear matter power spectrum}",
    journal = {Monthly Notices of the Royal Astronomical Society},
    volume = {484},
    number = {4},
    pages = {5509-5529},
    year = {2019},
    month = {01},
    issn = {0035-8711},
    doi = {10.1093/mnras/stz197},
    url = {https://doi.org/10.1093/mnras/stz197},
    eprint = {https://academic.oup.com/mnras/article-pdf/484/4/5509/27790453/stz197.pdf},
}

@Article{Knabenhans2021,
  author    = {{Euclid Collaboration} and {Knabenhans}, M. and {Stadel et al.}, J.},
  journal   = {Monthly Notices of the Royal Astronomical Society},
  title     = "{Euclid preparation: IX. EuclidEmulator2 – power spectrum emulation with massive neutrinos and self-consistent dark energy perturbations}",
  year      = {2021},
  issn      = {1365-2966},
  month     = may,
  number    = {2},
  pages     = {2840--2869},
  volume    = {505},
  doi       = {10.1093/mnras/stab1366},
  publisher = {Oxford University Press (OUP)},
}

@Article{Valcin2019,
  author    = {Valcin, David and Villaescusa-Navarro, Francisco and Verde, Licia and Raccanelli, Alvise},
  journal   = {Journal of Cosmology and Astroparticle Physics},
  title     = "{BE-HaPPY: bias emulator for halo power spectrum including massive neutrinos}",
  year      = {2019},
  issn      = {1475-7516},
  month     = dec,
  number    = {12},
  pages     = {057--057},
  volume    = {2019},
  doi       = {10.1088/1475-7516/2019/12/057},
  publisher = {IOP Publishing},
}

@Article{Nishimichi2019,
  author    = {Nishimichi, Takahiro and Takada, Masahiro and Takahashi, Ryuichi and Osato, Ken and Shirasaki, Masato and Oogi, Taira and Miyatake, Hironao and Oguri, Masamune and Murata, Ryoma and Kobayashi, Yosuke and Yoshida, Naoki},
  journal   = {The Astrophysical Journal},
  title     = "{Dark Quest. I. Fast and Accurate Emulation of Halo Clustering Statistics and Its Application to Galaxy Clustering}",
  year      = {2019},
  issn      = {1538-4357},
  month     = oct,
  number    = {1},
  pages     = {29},
  volume    = {884},
  doi       = {10.3847/1538-4357/ab3719},
  publisher = {American Astronomical Society},
}

@ARTICLE{Ho2023,
       author = {{Ho}, Ming-Feng and {Bird}, Simeon and {Fernandez}, Martin A. and {Shelton}, Christian R.},
        title = "{MF-Box: multifidelity and multiscale emulation for the matter power spectrum}",
      journal = {MNRAS},
         year = 2023,
        month = dec,
       volume = {526},
       number = {2},
        pages = {2903-2919},
          doi = {10.1093/mnras/stad2901},
archivePrefix = {arXiv},
       eprint = {2306.03144},
 primaryClass = {astro-ph.CO},
       adsurl = {https://ui.adsabs.harvard.edu/abs/2023MNRAS.526.2903H}
}

@ARTICLE{Ho2022,
       author = {{Ho}, Ming-Feng and {Bird}, Simeon and {Shelton}, Christian R.},
        title = "{Multifidelity emulation for the matter power spectrum using Gaussian processes}",
      journal = {MNRAS},
         year = 2022,
        month = jan,
       volume = {509},
       number = {2},
        pages = {2551-2565},
          doi = {10.1093/mnras/stab3114},
archivePrefix = {arXiv},
       eprint = {2105.01081},
 primaryClass = {astro-ph.CO},
       adsurl = {https://ui.adsabs.harvard.edu/abs/2022MNRAS.509.2551H}
}

@software{Feng2018,
  author       = {Yu Feng and
                  Simeon Bird and
                  Lauren Anderson and
                  Andreu Font-Ribera and
                  Chris Pedersen},
  title        = "{MP-Gadget/MP-Gadget: A tag for getting a DOI}",
  month        = oct,
  year         = 2018,
  publisher    = {Zenodo},
  version      = {FirstDOI},
  doi          = {10.5281/zenodo.1451799},
  url          = {https://doi.org/10.5281/zenodo.1451799}
}

@article{Qian2012,
author = {Peter Z. G. Qian},
title = "{Sliced Latin Hypercube Designs}",
journal = {Journal of the American Statistical Association},
volume = {107},
number = {497},
pages = {393-399},
year = {2012},
publisher = {Taylor & Francis},
doi = {10.1080/01621459.2011.644132},


URL = { 
    
        https://doi.org/10.1080/01621459.2011.644132
    
    

},
eprint = { 
    
        https://doi.org/10.1080/01621459.2011.644132
    
    

}
}

@ARTICLE{Bird2023,
       author = {{Bird}, Simeon and {Fernandez}, Martin and {Ho}, Ming-Feng and {Qezlou}, Mahdi and {Monadi}, Reza and {Ni}, Yueying and {Chen}, Nianyi and {Croft}, Rupert and {Di Matteo}, Tiziana},
        title = "{PRIYA: a new suite of Lyman-{\ensuremath{\alpha}} forest simulations for cosmology}",
      journal = {\jcap},
         year = 2023,
        month = oct,
       volume = {2023},
       number = {10},
          eid = {037},
        pages = {037},
          doi = {10.1088/1475-7516/2023/10/037},
archivePrefix = {arXiv},
       eprint = {2306.05471},
 primaryClass = {astro-ph.CO},
       adsurl = {https://ui.adsabs.harvard.edu/abs/2023JCAP...10..037B}
}

@ARTICLE{Arico2021,
       author = {{Aric{\`o}}, Giovanni and {Angulo}, Raul E. and {Contreras}, Sergio and {Ondaro-Mallea}, Lurdes and {Pellejero-Iba{\~n}ez}, Marcos and {Zennaro}, Matteo},
        title = "{The BACCO simulation project: a baryonification emulator with neural networks}",
      journal = {\mnras},
         year = 2021,
        month = sep,
       volume = {506},
       number = {3},
        pages = {4070-4082},
          doi = {10.1093/mnras/stab1911},
archivePrefix = {arXiv},
       eprint = {2011.15018},
 primaryClass = {astro-ph.CO},
       adsurl = {https://ui.adsabs.harvard.edu/abs/2021MNRAS.506.4070A}
}

@ARTICLE{Villaescusa2020,
       author = {{Villaescusa-Navarro}, Francisco and {Hahn}, ChangHoon and {Massara}, Elena and {Banerjee}, Arka and {Delgado}, Ana Maria and {Ramanah}, Doogesh Kodi and {Charnock}, Tom and {Giusarma}, Elena and {Li}, Yin and {Allys}, Erwan and {Brochard}, Antoine and {Uhlemann}, Cora and {Chiang}, Chi-Ting and {He}, Siyu and {Pisani}, Alice and {Obuljen}, Andrej and {Feng}, Yu and {Castorina}, Emanuele and {Contardo}, Gabriella and {Kreisch}, Christina D. and {Nicola}, Andrina and {Alsing}, Justin and {Scoccimarro}, Roman and {Verde}, Licia and {Viel}, Matteo and {Ho}, Shirley and {Mallat}, Stephane and {Wandelt}, Benjamin and {Spergel}, David N.},
        title = "{The Quijote Simulations}",
      journal = {\apjs},
         year = 2020,
        month = sep,
       volume = {250},
       number = {1},
          eid = {2},
        pages = {2},
          doi = {10.3847/1538-4365/ab9d82},
archivePrefix = {arXiv},
       eprint = {1909.05273},
 primaryClass = {astro-ph.CO},
       adsurl = {https://ui.adsabs.harvard.edu/abs/2020ApJS..250....2V}
}

@ARTICLE{Auld2007,
       author = {{Auld}, T. and {Bridges}, M. and {Hobson}, M.~P. and {Gull}, S.~F.},
        title = "{Fast cosmological parameter estimation using neural networks}",
      journal = {\mnras},
         year = 2007,
        month = mar,
       volume = {376},
       number = {1},
        pages = {L11-L15},
          doi = {10.1111/j.1745-3933.2006.00276.x},
archivePrefix = {arXiv},
       eprint = {astro-ph/0608174},
 primaryClass = {astro-ph},
       adsurl = {https://ui.adsabs.harvard.edu/abs/2007MNRAS.376L..11A}
}

@ARTICLE{Auld2008,
       author = {{Auld}, T. and {Bridges}, M. and {Hobson}, M.~P.},
        title = "{COSMONET: fast cosmological parameter estimation in non-flat models using neural networks}",
      journal = {\mnras},
         year = 2008,
        month = jul,
       volume = {387},
       number = {4},
        pages = {1575-1582},
          doi = {10.1111/j.1365-2966.2008.13279.x},
archivePrefix = {arXiv},
       eprint = {astro-ph/0703445},
 primaryClass = {astro-ph},
       adsurl = {https://ui.adsabs.harvard.edu/abs/2008MNRAS.387.1575A}
}

@ARTICLE{Arico2021a,
       author = {{Aric{\`o}}, Giovanni and {Angulo}, Raul E. and {Zennaro}, Matteo},
        title = "{Accelerating Large-Scale-Structure data analyses by emulating Boltzmann solvers and Lagrangian Perturbation Theory}",
      journal = {arXiv e-prints},
         year = 2021,
        month = apr,
          eid = {arXiv:2104.14568},
        pages = {arXiv:2104.14568},
          doi = {10.48550/arXiv.2104.14568},
archivePrefix = {arXiv},
       eprint = {2104.14568},
 primaryClass = {astro-ph.CO},
       adsurl = {https://ui.adsabs.harvard.edu/abs/2021arXiv210414568A}
}

@ARTICLE{Spurio2022,
       author = {{Spurio Mancini}, Alessio and {Piras}, Davide and {Alsing}, Justin and {Joachimi}, Benjamin and {Hobson}, Michael P.},
        title = "{COSMOPOWER: emulating cosmological power spectra for accelerated Bayesian inference from next-generation surveys}",
      journal = {\mnras},
         year = 2022,
        month = apr,
       volume = {511},
       number = {2},
        pages = {1771-1788},
          doi = {10.1093/mnras/stac064},
archivePrefix = {arXiv},
       eprint = {2106.03846},
 primaryClass = {astro-ph.CO},
       adsurl = {https://ui.adsabs.harvard.edu/abs/2022MNRAS.511.1771S}
}

@ARTICLE{Nygaard2023,
       author = {{Nygaard}, Andreas and {Holm}, Emil Brinch and {Hannestad}, Steen and {Tram}, Thomas},
        title = "{CONNECT: a neural network based framework for emulating cosmological observables and cosmological parameter inference}",
      journal = {\jcap},
         year = 2023,
        month = may,
       volume = {2023},
       number = {5},
          eid = {025},
        pages = {025},
          doi = {10.1088/1475-7516/2023/05/025},
archivePrefix = {arXiv},
       eprint = {2205.15726},
 primaryClass = {astro-ph.IM},
       adsurl = {https://ui.adsabs.harvard.edu/abs/2023JCAP...05..025N}
}

@ARTICLE{Gunther2022,
       author = {{G{\"u}nther}, Sven and {Lesgourgues}, Julien and {Samaras}, Georgios and {Sch{\"o}neberg}, Nils and {Stadtmann}, Florian and {Fidler}, Christian and {Torrado}, Jes{\'u}s},
        title = "{CosmicNet II: emulating extended cosmologies with efficient and accurate neural networks}",
      journal = {\jcap},
         year = 2022,
        month = nov,
       volume = {2022},
       number = {11},
          eid = {035},
        pages = {035},
          doi = {10.1088/1475-7516/2022/11/035},
archivePrefix = {arXiv},
       eprint = {2207.05707},
 primaryClass = {astro-ph.CO},
       adsurl = {https://ui.adsabs.harvard.edu/abs/2022JCAP...11..035G}
}

@ARTICLE{Bocquet2020,
       author = {{Bocquet}, Sebastian and {Heitmann}, Katrin and {Habib}, Salman and {Lawrence}, Earl and {Uram}, Thomas and {Frontiere}, Nicholas and {Pope}, Adrian and {Finkel}, Hal},
        title = "{The Mira-Titan Universe. III. Emulation of the Halo Mass Function}",
      journal = {\apj},
         year = 2020,
        month = sep,
       volume = {901},
       number = {1},
          eid = {5},
        pages = {5},
          doi = {10.3847/1538-4357/abac5c},
archivePrefix = {arXiv},
       eprint = {2003.12116},
 primaryClass = {astro-ph.CO},
       adsurl = {https://ui.adsabs.harvard.edu/abs/2020ApJ...901....5B}
}

@ARTICLE{Kwan2023,
       author = {{Kwan}, Juliana and {Saito}, Shun and {Leauthaud}, Alexie and {Heitmann}, Katrin and {Habib}, Salman and {Frontiere}, Nicholas and {Guo}, Hong and {Huang}, Song and {Pope}, Adrian and {Rodrigu{\'e}z-Torres}, Sergio},
        title = "{Galaxy Clustering in the Mira-Titan Universe. I. Emulators for the Redshift Space Galaxy Correlation Function and Galaxy-Galaxy Lensing}",
      journal = {\apj},
         year = 2023,
        month = jul,
       volume = {952},
       number = {1},
          eid = {80},
        pages = {80},
          doi = {10.3847/1538-4357/acd92f},
archivePrefix = {arXiv},
       eprint = {2302.12379},
 primaryClass = {astro-ph.CO},
       adsurl = {https://ui.adsabs.harvard.edu/abs/2023ApJ...952...80K}
}

@ARTICLE{Moran2023,
       author = {{Moran}, Kelly R. and {Heitmann}, Katrin and {Lawrence}, Earl and {Habib}, Salman and {Bingham}, Derek and {Upadhye}, Amol and {Kwan}, Juliana and {Higdon}, David and {Payne}, Richard},
        title = "{The Mira-Titan Universe - IV. High-precision power spectrum emulation}",
      journal = {\mnras},
         year = 2023,
        month = apr,
       volume = {520},
       number = {3},
        pages = {3443-3458},
          doi = {10.1093/mnras/stac3452},
archivePrefix = {arXiv},
       eprint = {2207.12345},
 primaryClass = {astro-ph.CO},
       adsurl = {https://ui.adsabs.harvard.edu/abs/2023MNRAS.520.3443M}
}

@ARTICLE{Takada2014,
       author = {{Takada}, Masahiro and {Ellis}, Richard S. and {Chiba}, Masashi and {Greene}, Jenny E. and {Aihara}, Hiroaki and {Arimoto}, Nobuo and {Bundy}, Kevin and {Cohen}, Judith and {Dor{\'e}}, Olivier and {Graves}, Genevieve and {Gunn}, James E. and {Heckman}, Timothy and {Hirata}, Christopher M. and {Ho}, Paul and {Kneib}, Jean-Paul and {Le F{\`e}vre}, Olivier and {Lin}, Lihwai and {More}, Surhud and {Murayama}, Hitoshi and {Nagao}, Tohru and {Ouchi}, Masami and {Seiffert}, Michael and {Silverman}, John D. and {Sodr{\'e}}, Laerte and {Spergel}, David N. and {Strauss}, Michael A. and {Sugai}, Hajime and {Suto}, Yasushi and {Takami}, Hideki and {Wyse}, Rosemary},
        title = "{Extragalactic science, cosmology, and Galactic archaeology with the Subaru Prime Focus Spectrograph}",
      journal = {\pasj},
         year = 2014,
        month = feb,
       volume = {66},
       number = {1},
          eid = {R1},
        pages = {R1},
          doi = {10.1093/pasj/pst019},
archivePrefix = {arXiv},
       eprint = {1206.0737},
 primaryClass = {astro-ph.CO},
       adsurl = {https://ui.adsabs.harvard.edu/abs/2014PASJ...66R...1T}
}

@ARTICLE{Casares2024,
       author = {{S{\'a}ez-Casares}, I. and {Rasera}, Y. and {Richardson}, T.~R.~G. and {Corasaniti}, P. -S.},
        title = "{The e-MANTIS emulator: Fast and accurate predictions of the halo mass function in f(R)CDM and wCDM cosmologies}",
      journal = {\aap},
         year = 2024,
        month = nov,
       volume = {691},
          eid = {A323},
        pages = {A323},
          doi = {10.1051/0004-6361/202450193},
archivePrefix = {arXiv},
       eprint = {2410.05226},
 primaryClass = {astro-ph.CO},
       adsurl = {https://ui.adsabs.harvard.edu/abs/2024A&A...691A.323S}
}

@ARTICLE{Bonici2025,
       author = {{Bonici}, Marco and {D'Amico}, Guido and {Bel}, Julien and {Carbone}, Carmelita},
        title = "{Effort: a fast and differentiable emulator for the Effective Field Theory of the Large Scale Structure of the Universe}",
      journal = {arXiv e-prints},
         year = 2025,
        month = jan,
          eid = {arXiv:2501.04639},
        pages = {arXiv:2501.04639},
          doi = {10.48550/arXiv.2501.04639},
archivePrefix = {arXiv},
       eprint = {2501.04639},
 primaryClass = {astro-ph.CO},
       adsurl = {https://ui.adsabs.harvard.edu/abs/2025arXiv250104639B}
}

@ARTICLE{Bonici2024,
       author = {{Bonici}, Marco and {Bianchini}, Federico and {Ruiz-Zapatero}, Jaime},
        title = "{Capse.jl: efficient and auto-differentiable CMB power spectra emulation}",
      journal = {The Open Journal of Astrophysics},
         year = 2024,
        month = jan,
       volume = {7},
          eid = {10},
        pages = {10},
          doi = {10.21105/astro.2307.14339},
archivePrefix = {arXiv},
       eprint = {2307.14339},
 primaryClass = {astro-ph.CO},
       adsurl = {https://ui.adsabs.harvard.edu/abs/2024OJAp....7E..10B}
}

@ARTICLE{Chen2025,
       author = {{Chen}, Zhao and {Yu}, Yu and {Han}, Jiaxin and {Jing}, Y.~P.},
        title = "{CSST Cosmological Emulator I: Matter Power Spectrum Emulation with one percent accuracy}",
      journal = {arXiv e-prints},
         year = 2025,
        month = feb,
          eid = {arXiv:2502.11160},
        pages = {arXiv:2502.11160},
          doi = {10.48550/arXiv.2502.11160},
archivePrefix = {arXiv},
       eprint = {2502.11160},
 primaryClass = {astro-ph.CO},
       adsurl = {https://ui.adsabs.harvard.edu/abs/2025arXiv250211160C}
}

@ARTICLE{Gong2019,
       author = {{Gong}, Yan and {Liu}, Xiangkun and {Cao}, Ye and {Chen}, Xuelei and {Fan}, Zuhui and {Li}, Ran and {Li}, Xiao-Dong and {Li}, Zhigang and {Zhang}, Xin and {Zhan}, Hu},
        title = "{Cosmology from the Chinese Space Station Optical Survey (CSS-OS)}",
      journal = {\apj},
         year = 2019,
        month = oct,
       volume = {883},
       number = {2},
          eid = {203},
        pages = {203},
          doi = {10.3847/1538-4357/ab391e},
archivePrefix = {arXiv},
       eprint = {1901.04634},
 primaryClass = {astro-ph.CO},
       adsurl = {https://ui.adsabs.harvard.edu/abs/2019ApJ...883..203G}
}

@misc{T2N2025,
  author       = {{Yang}, Yanhui and {Bird}, Simeon and {Ho}, Ming-Feng and {Qezlou}, Mahdi},
  title        = "{T2N-MusE: Triple-2 Neural Network Multifidelity Cosmological Emulation Framework}",
  year         = {2025},
  url          = {https://github.com/astro-YYH/T2N-MusE},
  note         = "{GitHub repository}"
}

@ARTICLE{Yang2025,
       author = {{Yang}, Yanhui and {Bird}, Simeon and {Ho}, Ming-Feng},
        title = "{Ten-parameter simulation suite for cosmological emulation beyond {\ensuremath{\Lambda}}CDM}",
      journal = {\prd},
         year = 2025,
        month = apr,
       volume = {111},
       number = {8},
          eid = {083529},
        pages = {083529},
          doi = {10.1103/PhysRevD.111.083529},
archivePrefix = {arXiv},
       eprint = {2501.06296},
 primaryClass = {astro-ph.CO},
       adsurl = {https://ui.adsabs.harvard.edu/abs/2025PhRvD.111h3529Y}
}

@ARTICLE{Cabayol-Garcia2023,
       author = {{Cabayol-Garcia}, L. and {Chaves-Montero}, J. and {Font-Ribera}, A. and {Pedersen}, C.},
        title = "{A neural network emulator for the Lyman-{\ensuremath{\alpha}} forest 1D flux power spectrum}",
      journal = {\mnras},
         year = 2023,
        month = nov,
       volume = {525},
       number = {3},
        pages = {3499-3515},
          doi = {10.1093/mnras/stad2512},
archivePrefix = {arXiv},
       eprint = {2305.19064},
 primaryClass = {astro-ph.CO},
       adsurl = {https://ui.adsabs.harvard.edu/abs/2023MNRAS.525.3499C}
}

@ARTICLE{Diao2025,
       author = {{Diao}, Kangning and {Mao}, Yi},
        title = "{Multi-fidelity emulator for large-scale 21 cm lightcone images: a few-shot transfer learning approach with generative adversarial network}",
      journal = {arXiv e-prints},
         year = 2025,
        month = feb,
          eid = {arXiv:2502.04246},
        pages = {arXiv:2502.04246},
          doi = {10.48550/arXiv.2502.04246},
archivePrefix = {arXiv},
       eprint = {2502.04246},
 primaryClass = {astro-ph.IM},
       adsurl = {https://ui.adsabs.harvard.edu/abs/2025arXiv250204246D}
}

@ARTICLE{Loshchilov2017,
       author = {{Loshchilov}, Ilya and {Hutter}, Frank},
        title = "{Decoupled Weight Decay Regularization}",
      journal = {arXiv e-prints},
         year = 2017,
        month = nov,
          eid = {arXiv:1711.05101},
        pages = {arXiv:1711.05101},
          doi = {10.48550/arXiv.1711.05101},
archivePrefix = {arXiv},
       eprint = {1711.05101},
 primaryClass = {cs.LG},
       adsurl = {https://ui.adsabs.harvard.edu/abs/2017arXiv171105101L}
}

@ARTICLE{Kingma2014,
       author = {{Kingma}, Diederik P. and {Ba}, Jimmy},
        title = "{Adam: A Method for Stochastic Optimization}",
      journal = {arXiv e-prints},
         year = 2014,
        month = dec,
          eid = {arXiv:1412.6980},
        pages = {arXiv:1412.6980},
          doi = {10.48550/arXiv.1412.6980},
archivePrefix = {arXiv},
       eprint = {1412.6980},
 primaryClass = {cs.LG},
       adsurl = {https://ui.adsabs.harvard.edu/abs/2014arXiv1412.6980K}
}

@ARTICLE{Ramachandran2017,
       author = {{Ramachandran}, Prajit and {Zoph}, Barret and {Le}, Quoc V.},
        title = "{Searching for Activation Functions}",
      journal = {arXiv e-prints},
         year = 2017,
        month = oct,
          eid = {arXiv:1710.05941},
        pages = {arXiv:1710.05941},
          doi = {10.48550/arXiv.1710.05941},
archivePrefix = {arXiv},
       eprint = {1710.05941},
 primaryClass = {cs.NE},
       adsurl = {https://ui.adsabs.harvard.edu/abs/2017arXiv171005941R}
}

@ARTICLE{Paszke2019,
       author = {{Paszke}, Adam and {Gross}, Sam and {Massa}, Francisco and {Lerer}, Adam and {Bradbury}, James and {Chanan}, Gregory and {Killeen}, Trevor and {Lin}, Zeming and {Gimelshein}, Natalia and {Antiga}, Luca and {Desmaison}, Alban and {K{\"o}pf}, Andreas and {Yang}, Edward and {DeVito}, Zach and {Raison}, Martin and {Tejani}, Alykhan and {Chilamkurthy}, Sasank and {Steiner}, Benoit and {Fang}, Lu and {Bai}, Junjie and {Chintala}, Soumith},
        title = "{PyTorch: An Imperative Style, High-Performance Deep Learning Library}",
      journal = {arXiv e-prints},
         year = 2019,
        month = dec,
          eid = {arXiv:1912.01703},
        pages = {arXiv:1912.01703},
          doi = {10.48550/arXiv.1912.01703},
archivePrefix = {arXiv},
       eprint = {1912.01703},
 primaryClass = {cs.LG},
       adsurl = {https://ui.adsabs.harvard.edu/abs/2019arXiv191201703P}
}

@InProceedings{Bergstra2013,
  title = 	 {Making a Science of Model Search: Hyperparameter Optimization in Hundreds of Dimensions for Vision Architectures},
  author = 	 {Bergstra, James and Yamins, Daniel and Cox, David},
  booktitle = 	 {Proceedings of the 30th International Conference on Machine Learning},
  pages = 	 {115--123},
  year = 	 {2013},
  editor = 	 {Dasgupta, Sanjoy and McAllester, David},
  volume = 	 {28},
  number =       {1},
  series = 	 {Proceedings of Machine Learning Research},
  address = 	 {Atlanta, Georgia, USA},
  month = 	 {17--19 Jun},
  publisher =    {PMLR},
  url = 	 {https://proceedings.mlr.press/v28/bergstra13.html}
}

@inproceedings{Kohavi1995,
  author = {Kohavi, Ron},
  booktitle = {Proceedings of the {I}nternational {J}oint {C}onference on {A}rtificial
	{I}ntelligence ({IJCAI})},
  pages = {1137--1143},
  publisher = {Morgan Kaufmann},
  title = {A study of cross-validation and {B}ootstrap for accuracy estimation
	and model selection},
  year = 1995
}

@ARTICLE{Yang2025b,
       author = {{Yang}, Yanhui and {Bird}, Simeon and {Ho}, Ming-Feng and {Qezlou}, Mahdi},
        title = "{Ten-dimensional neural network emulator for the nonlinear matter power spectrum}",
      journal = {arXiv e-prints},
         year = 2025,
        month = jul,
          eid = {arXiv:2507.07177},
        pages = {arXiv:2507.07177},
          doi = {10.48550/arXiv.2507.07177},
archivePrefix = {arXiv},
       eprint = {2507.07177},
 primaryClass = {astro-ph.CO},
       adsurl = {https://ui.adsabs.harvard.edu/abs/2025arXiv250707177Y}
}

@BOOK{2006gpml.book.....R,
       author = {{Rasmussen}, Carl Edward and {Williams}, Christopher K.~I.},
        title = "{Gaussian Processes for Machine Learning}",
         year = 2006,
       adsurl = {https://ui.adsabs.harvard.edu/abs/2006gpml.book.....R},
      publisher = {MIT Press}
}

@book{garnett_bayesoptbook_2023,
  author    = {{Garnett}, Roman},
  title     = "{{Bayesian Optimization}}",
  year      = {2023},
  publisher = {Cambridge University Press}
}

@ARTICLE{Hestness:2017arXiv171200409H,
       author = {{Hestness}, Joel and {Narang}, Sharan and {Ardalani}, Newsha and {Diamos}, Gregory and {Jun}, Heewoo and {Kianinejad}, Hassan and {Patwary}, Md. Mostofa Ali and {Yang}, Yang and {Zhou}, Yanqi},
        title = "{Deep Learning Scaling is Predictable, Empirically}",
      journal = {arXiv e-prints},
         year = 2017,
        month = dec,
          eid = {arXiv:1712.00409},
        pages = {arXiv:1712.00409},
          doi = {10.48550/arXiv.1712.00409},
archivePrefix = {arXiv},
       eprint = {1712.00409},
 primaryClass = {cs.LG},
       adsurl = {https://ui.adsabs.harvard.edu/abs/2017arXiv171200409H}
}

@article{scikit-learn,
  title={Scikit-learn: Machine Learning in {P}ython},
  author={Pedregosa, F. and Varoquaux, G. and Gramfort, A. and Michel, V.
          and Thirion, B. and Grisel, O. and Blondel, M. and Prettenhofer, P.
          and Weiss, R. and Dubourg, V. and Vanderplas, J. and Passos, A. and
          Cournapeau, D. and Brucher, M. and Perrot, M. and Duchesnay, E.},
  journal={Journal of Machine Learning Research},
  volume={12},
  pages={2825--2830},
  year={2011}
}

@ARTICLE{Zhang2025,
       author = {{Zhang}, Fan and {Luo}, Yifang and {Li}, Bohua and {Cao}, Ruihan and {Peng}, Wenjin and {Meyers}, Joel and {Shapiro}, Paul R.},
        title = "{SageNet: Fast Neural Network Emulation of the Stiff-amplified Gravitational Waves from Inflation}",
      journal = {arXiv e-prints},
         year = 2025,
        month = apr,
          eid = {arXiv:2504.04054},
        pages = {arXiv:2504.04054},
          doi = {10.48550/arXiv.2504.04054},
archivePrefix = {arXiv},
       eprint = {2504.04054},
 primaryClass = {astro-ph.CO},
       adsurl = {https://ui.adsabs.harvard.edu/abs/2025arXiv250404054Z}
}

\end{document}